\documentclass[12pt]{article}
\usepackage{rotating} 
\usepackage{adjustbox,lipsum}
\usepackage{float}
\usepackage{amssymb}
\usepackage{amsthm}
\usepackage{amsmath}
\usepackage{latexsym}
\usepackage[noend]{algpseudocode}
\usepackage{epsfig}
\usepackage{graphicx}
\usepackage{wasysym}
\usepackage{threeparttable}
\usepackage{natbib}
\usepackage{color}
\usepackage{algorithm}
\usepackage[noend]{algpseudocode}
\usepackage{epstopdf}
\usepackage{caption}
\usepackage{subcaption}
\usepackage{booktabs}

\newcommand{\ind}{\stackrel{\mathrm{ind}}{\sim}}

\usepackage{multirow}
\usepackage[breaklinks]{hyperref}

\usepackage{enumitem}
\usepackage{caption}
\usepackage{algorithm}

\def\boxit#1{\vbox{\hrule\hbox{\vrule\kern6pt
          \vbox{\kern6pt#1\kern6pt}\kern6pt\vrule}\hrule}}

\def\bse{\begin{eqnarray*}}
\def\ese{\end{eqnarray*}}
\def\be{\begin{eqnarray}}
\def\ee{\end{eqnarray}}
\def\bq{\begin{equation}}
\def\eq{\end{equation}}
\def\bse{\begin{eqnarray*}}
\def\ese{\end{eqnarray*}}

\usepackage{mathptmx}      
\usepackage{bm}

 {\begin{list}{}%
         {\setlength{\leftmargin}{#1}}%
         \item[]%
 }
 {\end{list}}
\usepackage{setspace}

\def\bz{\textbf{z}}

\def\by{\textbf{y}}

\def\bfbeta{\bm{\beta}}

\begin{document}
\thispagestyle{empty} \baselineskip=28pt

\begin{center}
	{\LARGE{\bf Joint spatio-temporal analysis of multiple response types using the hierarchical generalized transformation model with application to coronavirus disease 2019 and social distancing}}
\end{center}

\baselineskip=12pt

\vskip 2mm
\begin{center}
Jonathan R. Bradley\footnote{(\baselineskip=10pt to whom correspondence should be addressed) Department of Statistics, Florida State University, 117 N. Woodward Ave., Tallahassee, FL 32306-4330, jrbradley@fsu.edu}
\end{center}
%
%
%
%
\vskip 4mm

\begin{center}
\large{{\bf Abstract}}
\end{center}
Social distancing can be described as an effort to maintain a physical distance between individuals and has become a necessary public health measure to combat cornoavirus disease 2019 (COVID-19). Social distancing is known to weaken incidences and deaths due to COVID-19, however, there are detrimental economic and psychological effects. This motivates us to analyze incidences (and deaths) of COVID-19 along with a measure of the health of the US economy (i.e., the adjusted closing price of the Dow Jones Industrial), and a measure of the public interest in COVID-19 through Google Trends data. The model we implement is developed to be easily adapted to a data scientist's preferred method for continuous data, which is done to aid future analyses of this important dataset. This dataset consists of multiple response types (e.g., continuous-valued, count-valued, binomial counts). Thus, we introduce a reasonable easy-to-implement all-purpose method that ``converts'' a statistical model for continuous responses (the preferred model) into a Bayesian model for multi-response data sets. To do this, we transform the data such that the continuous-valued transformed data can be reasonably modeled using the preferred model and the transformation itself is treated as unknown. The implementation of our approach involves two steps. The first step produces posterior replicates of the transformed data using a latent conjugate multivariate (LCM) model. The second step involves generating values from the posterior distribution implied by the preferred model. We refer to our model as the hierarchical generalized transformation (HGT) model. In a simulation, we demonstrate the flexibility of the HGT model by incorporating two different preferred models: Bayesian additive regression trees (BART) and the spatial mixed effects (spatio-temporal mixed effects) models. We provide a thorough joint multiple-response spatio-temporal analysis of COVID-19 cases, the adjust closing price of the Dow Jones Industrial, and Google Trends data.
\baselineskip=12pt

%
%
%

\baselineskip=12pt
\par\vfill\noindent
\par\medskip\noindent
\clearpage\pagebreak\newpage \pagenumbering{arabic}
\baselineskip=24pt
\section{Introduction} COVID-19 was first detected in a live animal market in Wuhan City within the Hubei Province of China. This virus spreads easily from person to person, and there are cases of this virus where an individual is unsure of how they became infected (i.e., community spread). To date, there is no vaccine to prevent COVID-19, which has become a pandemic. As such, many governmental organizations, including the Centers for Disease Control and Prevention (CDC), have advised placing distance between yourself and other individuals (i.e., social distancing). Social distancing is an important public health measure that reduces close contact with people that may be infected by maintaining physical distance between all individuals \citep{wilder2020isolation,zhang2020age}. However, social distancing comes as a cost, and can be detrimental to economies and cause psychological distress \citep{long2020social}. With the negative effects of COVID-19 and social distancing in mind, we are interested in performing a joint spatio-temporal analysis of reported deaths and cases of COVID-19, the daily adjusted closing price of the Dow Jones Industrial (DJI), and a Google Trends data on searches of ``coronavirus.''

The data on reported deaths and cases of COVID-19 were obtained from the Johns Hopkins University Center for Systems Science and Engineering (JHU CCSE) Coronavirus repository (publicly available at https://github.com/CSSEGISandData/COVID-19), a subset of which, is made available in the R package \texttt{coronavirus} \citep{covid19Pack}. Cases, recoveries and mortality counts are available over regions (i.e., country or province) and discrete time (daily). In this article, we model these counts using a Poisson distribution, and our main interest lies in estimating the mean number of reported deaths and cases of COVID-19, and estimating its dependence with interest in COVID-19 and DJI data.

The number of Google searches of ``coronavirus'' is indicative of the high interest on COVID-19 and can act as a loose proxy for the public interest in COVID-19. This search information is made available through Google Trends data \citep{google}. Google Trends provide daily time series of an ``interest'' measure of searches on Google. This interest measure is defined on a scale from zero to one hundred with 100 indicating high interest and zero indicating low interest. In this article, we model the Google Trends interest score for the search ``coronavirus'' as binomial with sample size 100, since this response is a non-negative, integer-valued response that is bounded above by 100. We are interested in estimating the mean interest measure and estimating its dependence on the reported deaths and cases of COVID-19 and DJI data.

The DJI follows 30 publicly owned blue chip (i.e., nationally recognized and financially secure) companies that trade on the New York Stock Exchange (NYSE) and the National Association of Securities Dealers Automated Quotations (NASDAQ). It is a benchmark for blue-chip stocks and is often treated as a measure of the economic health of the US. This data was obtained through Yahoo Finance \citep{yahoo}. We model the adjusted daily closing price with a Gaussian distribution, since it is continuous valued. Our main interest in DJI is in determining and summarizing the relationship between the adjusted closing price with both interest in COVID-19 and reported cases and deaths due to COVID-19.

A major difficulty in jointly analyzing these data is that the response types are different (i.e., Poisson, binomial, and Gaussian). There are several methods for jointly modeling data consisting of multiple response types, however these approaches often require substantial methodological development, or creates clear computational difficulties. For example, Markov models \citet{allen}, copulas \citep{rank2,rank1, copula1,copula2}, multi-task learning models \citep{Argyriou,Kim,Yang}, regression trees, and random forests \citep{htf,forest} have been adapted to this multiple response setting. However, these methods do not immediately incorporate a data scientist's preferred model. An important goal of this article is to allow our model to be flexible enough that it can be adapted to other data scientist's preferred model. There has been a call to action for researchers to analyze COVID-19 \citep{OSTP}, and because of this, it is desirable to have tool that makes it easy for data scientists to jointly analyze Google Trends, DJI, and incidences of COVID-19 using their preferred model. While our proposed model allows for this flexibility, it can interpreted as a simple combination of two existing methods: generalized linear mixed effects models \citep[e.g., see][for a standard reference]{glm-mcculloch} and LCMs \citep{bradleyLCM}.

The GLMM is a standard approach to model non-Gaussian data. For example, Bernoulli data is modeled hierarchically, where the logit of the probability of success can be analyzed using a data scientist's preferred model. GLMMs lack conjugacy, which creates noticeable difficulty when implementing a GLMM on a modern high-dimensional data set. A more recent alternative is the LCM. Basic theoretical results and empirical analyses in \citet{bradleyPMSTM}, \citet{hu2018bayesian}, \citet{yang2019bayesian}, \citet{bradley2019spatio}, and \citet{bradleyLCM} suggest that one can outperform a standard GLMM (specifically Latent Gaussian Process (LGP) models) in terms of prediction error. However, both the GLMM and LCM requires the preferred model to be a mixed effects model, and the LCM requires one to modify the distribution of random effects to follow the appropriate distribution based on conjugacy.

A classical approach is to \textit{transform} the data, so that the transformed data can be reasonably modeled using the distribution assumed by the preferred model. In the non-Bayesian settings this literature is extremely well-developed and includes the Box-Cox transformations \citep{box1964analysis}, the alternating conditional expectations \citep[ACE;][]{breiman1985estimating} algorithm, graphical techniques \citep{mcculloch1993fitting}, and the Yeo-Johnson power transformation \citep{yeo2000new}, among other techniques. More recently developments in rank based algorithms \citep{servin2007imputation,mccaw2019omnibus,beasley2009rank} and quantile-matching \citep{mccullagh2020likelihood} have also been proposed in the non-Bayesian setting. It is important to note that Bayesian models for transformations have been proposed as well, but focus on the case where continuous non-normal data are observed and the preferred model assumes normality. In particular, these Bayesian models put a prior on the free parameter within the Box-Cox transformation or the Yeo-Johnson power transformation \citep{kim2013bayesian,charitidou2015bayesian,charitidou2018objective}. No such Bayesian model has been developed to analyze multi-response response data using any preferred model for a continuous response.

There are three distributions that define our hierarchical generalized transformation (HGT) model: (a) the distribution of the data given a transformation, (b) the prior distribution of the transformation, and (c) the distribution of the process of interest (i.e., the aforementioned preferred model). In this article, we model the data given a transformation (a) using members from the exponential family. Specifically, given a transformation, continuous data follows the normal distribution, categorical data follows the binomial distribution, and count-data follow the Poisson distribution. These distributions are conjugate with the normal, the logit-beta \citep{gao2019bayesian,bradley2019spatio} and the log-gamma distributions \citep{bradleyPMSTM,hu2018bayesian,bradleyLCM,yang2019bayesian}, which are special cases of the Diaconis-Ylvishaker (DY) distribution \citep[e.g., see][for key references]{diaconis,chen2003conjugate}. Consequently, the prior distribution of the transformation (b) is modeled with a DY distribution, which defines an LCM model for the transformations.

 While we are motivated by COVID-19 and the detrimental impacts of social distances, the methodology developed in this manuscript is of independent interest, since this is a new way in Bayesian statistics to model non-Gaussian processes using models for continuous data. Furthermore, our methodology also allows one to analyze a single non-Gaussian response type in a straightforward manner. That is, the implementation of our approach can be done using composite sampling. In particular, the first step is to sample from the posterior distribution of the transformation. Then the second step is to sample from the conditional distribution of the latent process of interest given the transformation. This conditional distribution is derived from the preferred model.
 
 The first step of the composite sampler is computationally straightforward because the DY distribution is conjugate (and easy to sample from) with the exponential family. Additionally, the first step of this algorithm is important for the purpose of analyzing multiple response types. Specifically, at the end of the first step we obtain a replicate from the posterior distribution of the transformation (which is continuous valued). Thus, the first step of the composite sampling algorithm ``transforms'' the multi-response data into a continuous-valued quantity appropriate for the preferred model.
 
Implementation of the preferred model is unchanged in the second step of our composite sampling algorithm. This is particularly noteworthy, as many of the Bayesian statistical models derived for Gaussian data are not immediately computationally efficient in the non-Gaussian data setting \citep[e.g., see][for examples in the spatial setting]{bradleyhierarchical, kang-cressie-2011,katzfuss2012}. This is because GLMMs in the non-Gaussian setting have full-conditional distributions that are not Gaussian, and can not be sampled from immediately. Bayesian methods that do not have easy to sample from full-conditional distributions require difficult to tune Metropolis-Hastings algorithms \citep[e.g., see][for an example]{bradleyLCM}, inefficient rejection samplers \citep[e.g., see][]{Damien}, or significant reparameterization to make approximate Bayesian methods (that are only appropriate for small parameter spaces) practical \citep{rue,ne2}. The second step of our composite sampling algorithm allows one to circumvent this issue entirely, and simply use the computational strategies that were developed for the preferred model.

The two steps of our composite sampler can be seen as sequential smoothing. By ``smoothing'' we mean a function of the data that attempts to discover important features in the data \citep[e.g., see][for a standard reference]{simonoff2012smoothing}. Multiple layers of smoothing may lead to estimates that are ``oversmooth,'' in the sense that many features of the data are not captured. To avoid oversmoothing we specify the model so that the posterior distribution of the transformation is ``saturated.'' Recall a saturated model is one in which there exists at least as many parameters as there are data points, and fitting this model allows you to exactly recover the original data set. Hence, saturated models are often an extreme example of overfitting. Thus, in the first step of our composite sampler we choose to overfit the data, and in the second step we smooth overfitted values (again this is done to avoid oversmoothing). 

In the classical log-linear model literature, saturated models are useful for selecting more parsimonious models \citep[e.g., see][for a standard reference]{agresti2011categorical}. Specifically, the most parsimonious reduced model that is not significantly different (in terms of the deviance or chi-square statistic) from the saturated model is used for statistical inference. Consequently, specifying the transformation model to be saturated allows us to assess the goodness-of-fit of the preferred model in a fully Bayesian manner that is similar to what is done in classical residual analysis.

It has recently been shown that forecasts regarding COVID-19 requires sophisticated models. Following the results of \citet{donnat2020modeling}, we include spatio-temporal random effects through the use of basis function expansions \citep[e.g., see][for a standard reference]{cressie-wikle-book}. Additionally, to improve the performance of forecasting we adopt the training, validation, and testing data framework that has become standard among the machine learning literature \citep[e.g., see][for a standard reference]{htf}.

The remainder of this article is organized as follows. In Section~\ref{Section:2}, we introduce our motivating dataset and describe how standard modeling procedures are not appropriate for this dataset. Then, we introduce the HGT model to analyze multi-response data with unknown transformations in Section~\ref{Section:3}. Additionally, we provide a specific class of transformation models and an example model specification. Then in Section~\ref{Section:4}, we provide details on using training, validation, and testing data for statistical inference. A summary of all the Bayesian models used in our analysis is also provided. In Section~\ref{Section:5}, we give simulation studies to illustrate that our approach has been been developed in a manner that one can incorporate their preferred statistical model. In particular, we apply our approach to BART models and a spatio-temporal mixed effects (SME) model. Section~\ref{Section:6} contains our joint analysis of COVID-19 mortality, incidences and recoveries, along with Google Trends data, and DJI data. Section~\ref{Section:7} contains a discussion and derivations are provided in the appendices.

\begin{figure}[htp]
	\begin{center}
		\begin{tabular}{ccc}
			\hspace{-40pt}\includegraphics[width=6cm,height=6cm]{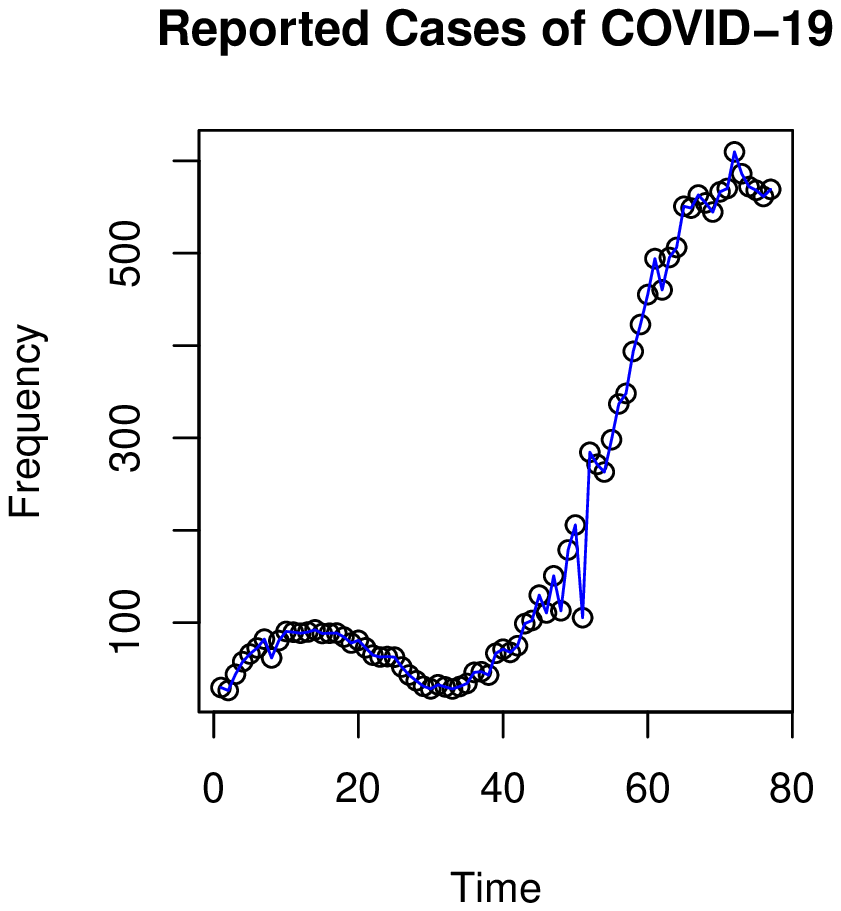}&\includegraphics[width=6cm,height=6cm]{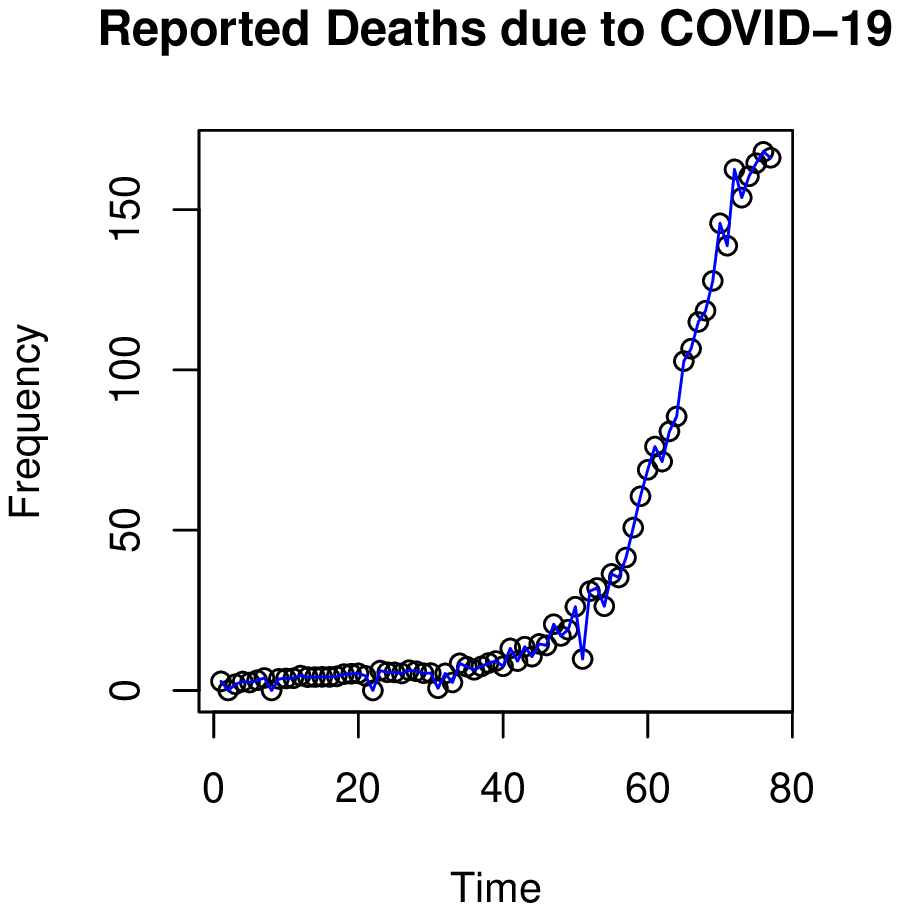}&\includegraphics[width=6cm,height=6cm]{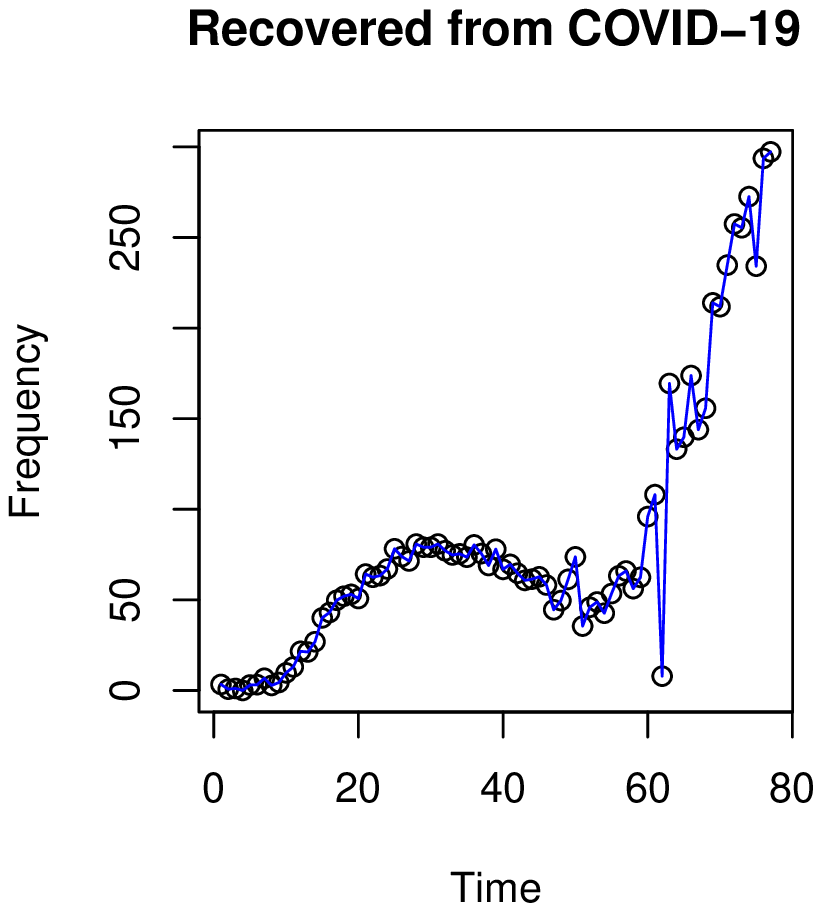}
		\end{tabular}
	\end{center}
\begin{center}
			\begin{tabular}{cc}
		\includegraphics[width=6cm,height=6cm]{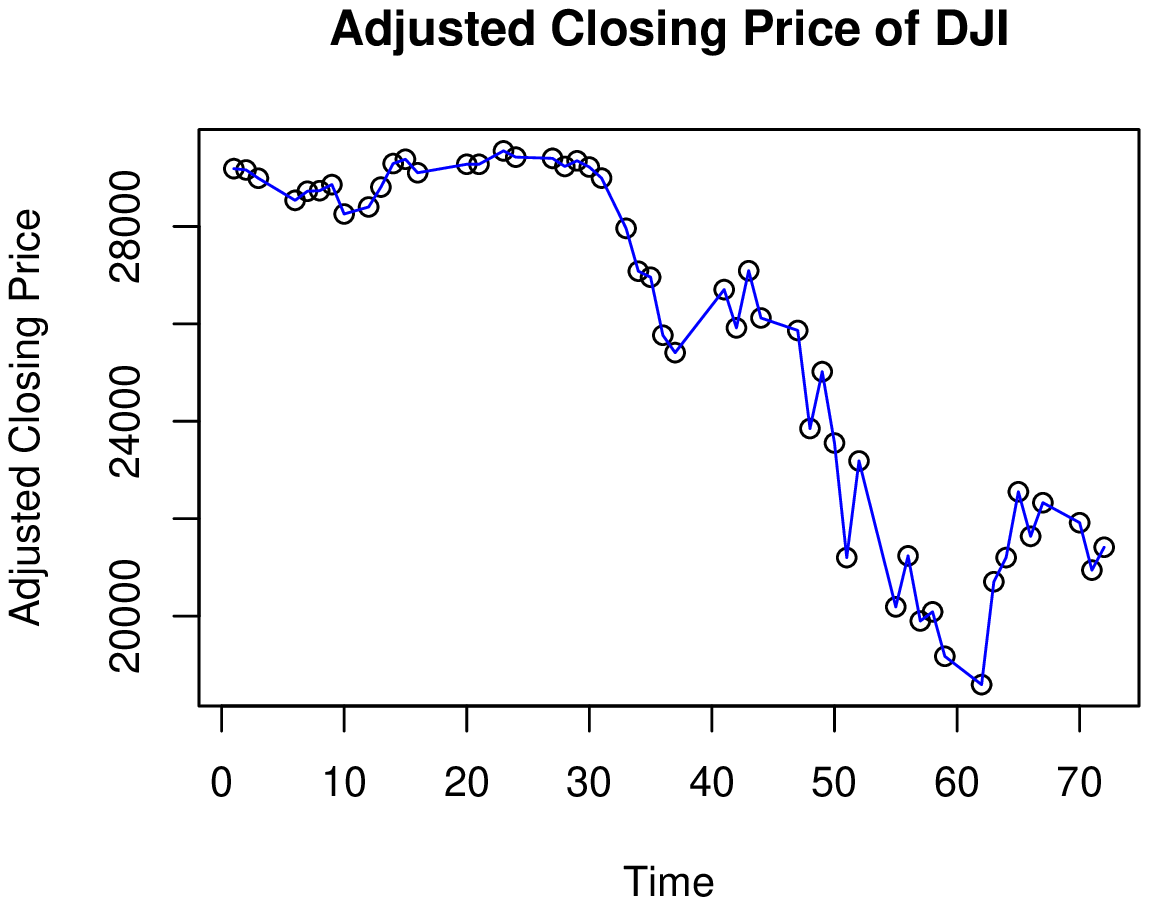}&\includegraphics[width=6cm,height=6cm]{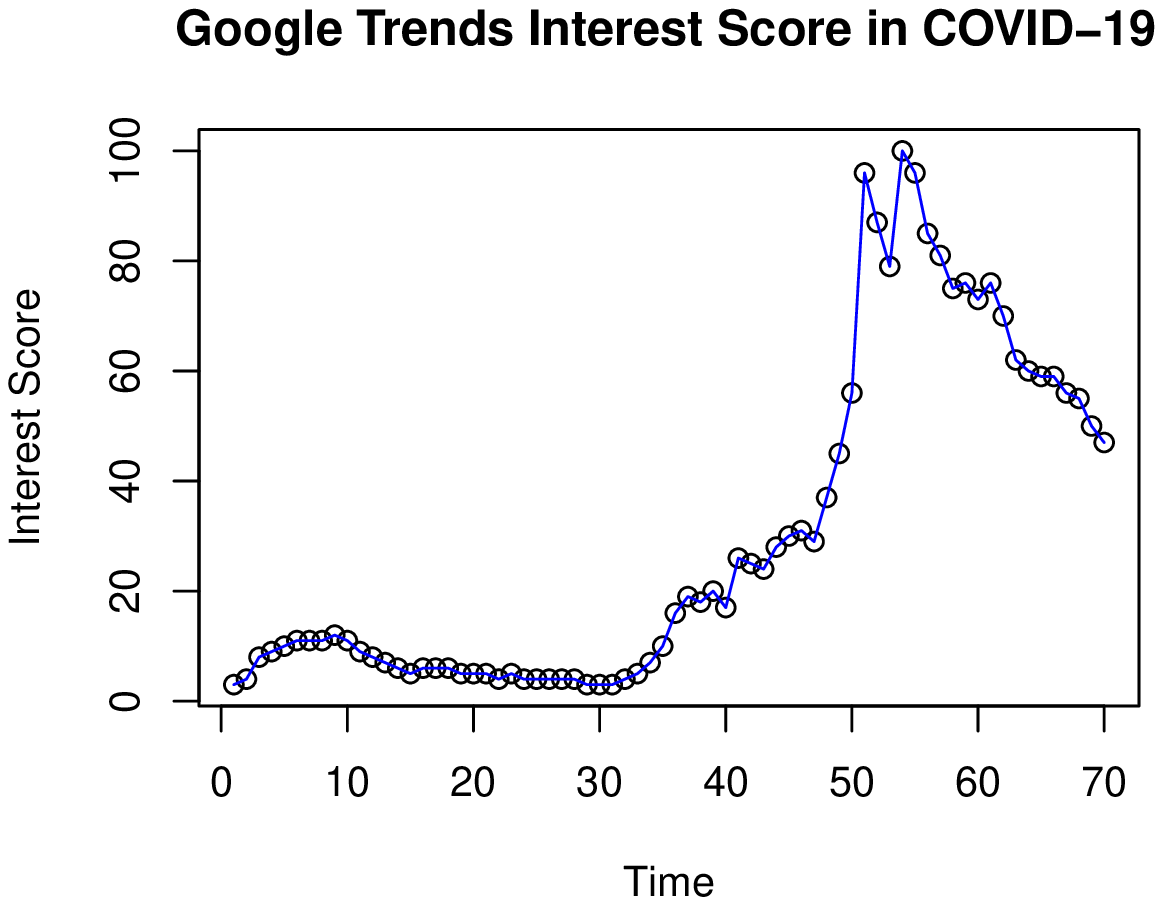}\\
	\end{tabular}
		\caption{We plot the number of reported COVID-19 infections (top left), reported COVID-19 deaths (top middle), the reported recoveries from COVID-19 (top right), the DJI adjusted closing price (bottom left), and the Google Trends interest score for searches of ``coronavirus'' (bottom right). Note that the DJI price data is not available on Saturday and Sundays. The black circles are the observed data, and blue lines connecting these points are added as a reference. The top row represents only a summary of available data, since we also observe these counts over 184 countries and 82 provinces.}\label{fig:data}
	\end{center}
\end{figure}

\section{Motivating Dataset}\label{Section:2} Denote the data with $Z_{ij}$, where $i$ indexes replicates and $j$ indexes response type such that $i = 1,\ldots, I_{j}$ and $j = 1, 2, 3$. We consider the setting where for each $i$, $Z_{i1}$ is continuous-valued, $Z_{i2}$ is integer-valued ranging from $0,\ldots, b_{i}$, and $Z_{i3}$ is count valued. Specifically, $Z_{i1}$ represents a measure of the adjusted closing price of DJI, $Z_{i2}$ is the integer-valued interest score for COVID-19 searches as computed by Google Trends (with $b_{i}\equiv 100$), and $i$ indexes the days ranging from January 22, 2020 to April 8, 2020. The data $Z_{i3}$ represents the $i$-th replicate of the number of COVID-19 cases, where for each $i$ there is an associated region (e.g., China) $A_{i} \subset \in [-180,180]\times [-90,90]$, day $t_{i}$ (between January 22, 2020 to April 8, 2020), and an indicator $d_{i}$ of whether or not the count consists of reported deaths. Let $d_{i} = 1$ if $Z_{i3}$ represents reported deaths and $d_{i} = 0$ otherwise. Likewise, let $u_{i}$ represent an indicator of whether or not the count consists of reported recoveries. Also let $t_{i} = 1,\dots, T=78$ represent each day between January 22, 2020 to April 8, 2020. In Figure \ref{fig:data}, we plot the number of reported COVID-19 infections, reported COVID-19 deaths, the DJI adjusted closing price, and the Google Trends interest score for searches of ``coronavirus.''

There are many ``off-the-shelf'' approaches that one might consider to analyze this data. For example, one might define the following linear model,
\begin{equation*}
Y_{1} = \textbf{x}_{i1}^{\prime}\bm{\beta}_{1} + \beta_{Y2}\sum_{j = 1}^{T}Y_{j2}I(t_{i} = j) + \beta_{Y3}\sum_{j = 1}^{T}Y_{j3}I(t_{i} = j) + \xi_{i1};\hspace{5pt} i = 1,\ldots, I_{1},
\end{equation*}
\noindent
where $\xi_{i1}$ is normally distributed with mean zero and variance $\sigma_{\xi}^{2}$, $I(\cdot)$ is an indicator function, $\beta_{Yk}\in \mathbb{R}$, $\bm{\beta}_{1}$ is an unknown $p$-dimensional vector, and $\textbf{x}_{i1}$ is a $p$-dimensional covariate vector. However, this conditionally specified model enforces a strong assumption of linearity between the different response types. Furthermore, the variability (and dependence) of $Y_{i2}$ and $Y_{i3}$ is ignored.

To incorporate the variability across response types (i.e., across $j$) and allow for non-linear relationships, one might also consider the following hierarchical model:
\begin{align}\label{glm}
\nonumber
Z_{i1} &\sim \mathrm{Normal}(Y_{i1},v)\\
\nonumber
Z_{i2} &\sim \mathrm{Binomial}\left\lbrace b_{i},\frac{\mathrm{exp}\left(Y_{i2}\right)}{1+\mathrm{exp}\left(Y_{i2}\right)}\right\rbrace\\
Z_{i3} &\sim \mathrm{Poisson}\left\lbrace\mathrm{exp}\left(Y_{ij}\right)\right\rbrace; \hspace{5pt}i = 1,\ldots, I_{j}, j = 1,2,3,
\end{align}
\noindent
where $Y_{ij}$ is an unobserved latent process, $\mathrm{Normal}(Y_{i1},v)$ is a shorthand for the normal distribution with mean $Y_{ij}\in \mathbb{R}$ and variance $v>0$, $\mathrm{Binomial}(b_{i},p)$ is a shorthand for the binomial distribution with $b_{i}>1$ number of trials and probability of success $p \in (0,1)$, and Poisson$(\mu_{ij})$ is a shorthand for the Poisson distribution with mean $\mu_{ij}$. The covariance between observations is determined by the model for $Y_{ij}$:
\begin{align}
\nonumber
cov(Z_{ij}, Z_{mk}) &= E\left\lbrace cov(Z_{ij}, Z_{mk})\vert Y_{ij},Y_{mk}\right\rbrace + cov\left\lbrace E(Z_{ij}\vert Y_{ij}), E(Z_{mk}\vert Y_{mk})\right\rbrace,\\
\label{covZZ}
&=  cov\left\lbrace E(Z_{ij}\vert Y_{ij}), E(Z_{mk}\vert Y_{mk})\right\rbrace=  cov\left\lbrace c_{ij}g_{j}^{-1}(Y_{ij}), c_{mk}g_{j}^{-1}(Y_{mk})\right\rbrace,
\end{align}
\noindent
for $i\ne m$ and $j\ne k$, where the functions $g_{1}(x_{i})= x_{i}$, $g_{2}(x_{i}) = log(x_{i}/1-x_{i})$, and $g_{3}(x_{i}) = log(x_{i})$ are referred to as ``link functions,'' and $c_{i1}=c_{i3} = 1$ and $c_{i2} = b_{i}$. Similarly, predicted values are determined by the model for $Y_{ij}$:
\begin{align}
\label{predZ}
E(Z_{ij}) &= E\left\lbrace E(Z_{ij}\vert Y_{ij})\right\rbrace = E\left\lbrace c_{ij}g_{j}^{-1}(Y_{ij})\right\rbrace.
\end{align}
\noindent
Thus, cross-dependence and predictions are modeled through the statistical model assumed for the process $Y_{ij}$, and a standard choice in this context is the GLMM:
\begin{equation}\label{glmm}
Y_{ij} = \textbf{x}_{ij}^{\prime}\bm{\beta}_{j} + \bm{S}_{ij}^{\prime}\bm{\eta} + \xi_{ij},
\end{equation}
where $\textbf{x}_{ij}$ is a known $p$-dimensional vector of covariates and $\textbf{S}_{ij}$ is a pre-specified $r$-dimensional vector of basis functions, $\bm{\beta}_{j} = (\beta_{1j},\ldots, \beta_{pj})^{\prime}$, $\bm{\eta} = (\eta_{1},\ldots, \eta_{r})^{\prime}$, $\beta_{kj}\ind \mathrm{Normal}(0,\sigma_{\beta}^{2})$, $\eta_{k}\ind \mathrm{Normal}(0,\sigma_{\eta}^{2})$, $\xi_{ij}\ind \mathrm{Normal}(0,\sigma_{\xi}^{2})$, $\sigma_{\beta}^{2}>0$, $\sigma_{\eta}^{2}>0$, and $\sigma_{\xi}^{2}>0$. Then, the cross-response spatio-temporal covariance implied by this model is $cov\left( Y_{ij}, Y_{mk}\vert \sigma_{\eta}^{2}\right)= \sigma_{\eta}^{2}\textbf{S}_{ij}^{\prime}\textbf{S}_{mk}$, which propagates through and enforces dependence in the data through Equation (\ref{covZZ}). The relationship between the different response types can be found by estimating the unknown function $\textbf{S}_{ij}^{\prime}\bm{\eta}$ (e.g., using posterior means and credible intervals).

Computationally, the GLMM is difficult to implement in a Bayesian context. For example, a Gibbs sampler requires one to simulate from the following full-conditional distributions \citep{gelfandGibbs}, and in this setting these distributions do not have a known form that is straightforward to simulate from. There are several approximate Bayesian computational tools available, however, for moderate sizes of $p$ and $r$ these approaches are not feasible. In particular, Hamiltonian Monte Carlo \citep[HMC; ][]{neal2011mcmc} and the integrated nested Laplace approximation \citep[INLA; ][]{rue} are only appropriate for small parameter spaces \citep[e.g,][suggests no more than 15 parameters when implementing INLA]{inla22}. Additionally, INLA only allows for marginal inference \citep{kristensen2015tmb}. The computational issues of the hierarchical model in (\ref{glm}) and (
\ref{glmm}) may become even more cumbersome when considering a different model for $Y_{ij}$. This is especially pertinent for our dataset, since the US government has put out a call to action \citep{OSTP} for data scientists to analyze COVID-19 datasets, and it would be preferable to have approach that is flexible enough for others to specify their own model for $Y_{ij}$ without major changes to implementation.

\section{The Hierarchical Generalized Transformation Model}\label{Section:3}

\subsection{Unknown Transformations of Multiple Response Types} \label{Section:3.1} One classical strategy to model non-Gaussian data is to impose a transformation such that,
\begin{equation}\label{preferred}
h_{j}(Z_{ij})\vert Y_{ij},\bm{\theta} \sim \mathrm{Dist}\left(Y_{ij},\bm{\theta}\right),\hspace{20pt} i = 1,\ldots, I_{j}, j = 1,2,3,
\end{equation}
\noindent
where $h(\cdot)$ is a transformation of the datum $Z_{ij}$, ``Dist'' is a short-hand used for a probability density function (pdf), $g_{j}\left\lbrace E(Z_{ij})\right\rbrace = Y_{ij}\in \mathbb{R}$ and  $\bm{\theta} \in \Theta \subset \mathbb{R}^{p}$. Additionally, $Y_{ij}$ is defined for $i = 1,\ldots, I$ and $j = 1,2, 3$, where $I \ge max(I_{1},I_{2},I_{3})$. Here, ``$\mathrm{Dist}\left(Y_{ij},\bm{\theta}\right)$'' represents the aforementioned preferred model. In what remains, inference on $\{Y_{ij}\}$ and $\bm{\theta}$ is the primary goal. To aid in our exposition we drop the functional notation for $h(\cdot)$ and write $h_{ij} = h_{j}(Z_{ij})$. As an example of ``Dist,'' suppose we assume
\begin{equation}\label{transformeddatamodel}
h_{ij} = Y_{ij}+\epsilon_{ij},
\end{equation}
\noindent
where $\epsilon_{ij}\ind \mathrm{Normal}(0,\sigma_{\epsilon}^{2})$ and $\sigma_{\epsilon}^{2}>0$, and the model on $Y_{ij}$ in (\ref{glmm}) is assumed.

Transformations convert a multiple response type data set (e.g., $\{Z_{ij}\}$) to a single response type data set (e.g., $\{h_{ij}\}$), since $h_{ij}$ follows a single distribution with a continuous support. Consequently, transformations have become a standard tool in analyzing multiple response types. Recall, transformations such as these have a long history including the box-cox transformations \citep{box1964analysis}, graphical techniques \citep{mcculloch1993fitting}, the alternating conditional expectations \citep[ACE;][]{breiman1985estimating} algorithm, and the Yeo-Johnson power transformation \citep[][among others]{yeo2000new}. 

In this paper, we introduce a Bayesian solution to the problem of an unknown transformation. In particular, we define pdf and probability mass functions (pmf), $f(Z_{ij}\vert h_{ij})$. We refer to these distributions as ``transformation models.'' In Section~\ref{Section:3.2}, we describe Bayesian implementation using a general transformation model and any well defined preferred model. Then, in Section~\ref{Section:3.3} the specification of the transformation model is given. Finally we provide an example in Section~\ref{Section:3.4}.

\subsection{General Bayesian Implementation}\label{Section:3.2} In this section, we describe Bayesian implementation of the model introduced in Section~\ref{Section:3.1}. Here, let $n = \sum_{j = 1}^{3}I_{j}$, the $n$-dimensional data vector $\textbf{z}_{trn} = \left(Z_{11},\ldots, Z_{I_{3}3}\right)^{\prime}$, the $n$-dimensional transformed data vector $\textbf{h} = \left(h_{11},\ldots, h_{I_{3}3}\right)^{\prime}$, $N = 3I \ge n$, and the $N$-dimensional latent process $\textbf{y} = \left(Y_{11},\ldots, Y_{I1},Y_{12},\ldots, Y_{I2},Y_{13},\ldots, Y_{I3}\right)^{\prime}$. Notice, that $I_{j}\le I$, which allows for missing values of $Y_{ij}$.

From (\ref{preferred}), the preferred model ``Dist'' is represented in terms of a hierarchical model:
\begin{align}\label{preferredHM}
\nonumber
&f(h_{ij}\vert Y_{ij}, \bm{\theta})m(\textbf{h}\vert \bm{\gamma}); \hspace{20pt}i = 1,\ldots, I_{j}, j = 1,2,3,\\
\nonumber
&f(\textbf{y}\vert \bm{\theta})\\
&f(\bm{\theta}), 
\end{align}
\noindent
where $m(\cdot)$ is a real-valued function of $\textbf{h}$, which we will define below. Following the terminology used in \citet{cressie-wikle-book}, we call $f(h_{ij}\vert Y_{ij}, \bm{\theta})m(\textbf{h}\vert \bm{\gamma})$ the ``transformed data model,'' $f(Y_{ij}\vert \bm{\theta})$ the ``process model,'' and $f(\bm{\theta})$ the ``parameter model'' (or prior). Bayes rule can be used to produce the following conditional distribution \citep[e.g., see][for a standard reference]{gelmanbook},
\begin{equation}\label{preferredModel}
f\left(\textbf{y}, \bm{\theta}\vert \textbf{h}\right) = \frac{f\left(\textbf{h}\vert \textbf{y}, \bm{\theta}\right) f (\textbf{y}\vert \bm{\theta}) f(\bm{\theta})}{\int \int f\left(\textbf{h}\vert \textbf{y}, \bm{\theta}\right) f (\textbf{y}\vert \bm{\theta}) f(\bm{\theta})d\textbf{y}d\bm{\theta} },
\end{equation}
\noindent
where we have assumed $h_{ij}$ is conditionally independent of $h_{km}$ given $Y_{ij}$ and $\bm{\theta}$ for $k\ne i$ and $m \ne j$ so that $f(\textbf{h}\vert \textbf{y}, \bm{\theta}) = \prod_{i}\prod_{j}f(h_{ij}\vert Y_{ij}, \bm{\theta})$. Similarly, one can use Bayes rule to produce the posterior distribution of the transformed data. That is,
\begin{equation}\label{posteriorTrans}
f(\textbf{h}\vert \textbf{z}_{trn}) = \frac{\int f(\textbf{z}_{trn}\vert \textbf{h})f(\textbf{h}\vert \bm{\gamma})f(\bm{\gamma})\hspace{2pt}d\bm{\gamma}}{\int \int f(\textbf{z}_{trn}\vert \textbf{h})f(\textbf{h}\vert \bm{\gamma}) f(\bm{\gamma})\hspace{2pt}d\textbf{h}\hspace{2pt}d\bm{\gamma}},
\end{equation}
\noindent
where the distribution $f(\textbf{h}\vert \bm{\gamma})$ is referred to as a ``transformation prior,'' the $q$-dimensional real-valued vector $\bm{\gamma}$ is referred to as a transformation hyperparameter, and the distribution $f(\bm{\gamma})$ is referred to as a ``transformation hyperprior.'' To guarantee that our choice of the transformation prior and transformed data model are consistent with each other we set $m(\textbf{h}\vert \bm{\gamma}) = f(\textbf{h}\vert \bm{\gamma})/\int \int f\left(\textbf{h}\vert \textbf{y}, \bm{\theta}\right) f (\textbf{y}\vert \bm{\theta}) f(\bm{\theta})d\textbf{y}d\bm{\theta}$.

Equations (\ref{preferredModel}) and (\ref{posteriorTrans}) can be used to produce a posterior distribution for $\textbf{y}$ and $\bm{\theta}$. That is, suppose $f(\textbf{h}\vert \textbf{y},\bm{\theta})$, $f( \textbf{y}\vert\bm{\theta})$, $f(\bm{\theta})$, $f(\textbf{z}_{trn}\vert \textbf{h})$,  $f(\textbf{h}\vert \bm{\gamma})$, and $f(\bm{\gamma})$ are proper. Suppose $\textbf{z}_{trn}$ is conditionally independent of $\bm{\gamma}$ given $\textbf{h}$, and $\textbf{z}_{trn}$ and $(\by^{\prime},\bm{\theta}^{\prime})^{\prime}$ are conditionally independent given $\textbf{h}$. Then:
\begin{align}
\label{bayesconverter}
f(\by,\bm{\theta}\vert \bz) &= \int f(\by,\bm{\theta}\vert \textbf{h})f(\textbf{h}\vert \bz) d\textbf{h}.
\end{align}
\noindent
The derivation of (\ref{bayesconverter}) can be found in Appendix A.

The model in (\ref{bayesconverter}) can easily be simulated from using a composite sampling scheme, provided that it is easy to simulate from $f(\by,\bm{\theta}\vert \textbf{h})$. Algorithm 1 gives the step-by-step implementation of how to simulate from the posterior distribution in (\ref{bayesconverter}). Here, we see that the implementation of the HGT model is similar to the bootstrap implementation, where we have replaced a resampling step with sampling from $f(\textbf{h}\vert \bz)$ and the full-conditional distributions associated with $\bm{\gamma}$. This similarity emphasizes the flexibility of allowing for unknown transformations in a Bayesian context, since the bootstrap algorithm is an established flexible approach in the literature \citep[e.g., see][for an early reference]{efron1992bootstrap}. Of course, the bootstrap algorithm produces replicates from a different distribution than that of Algorithm 1. Specifically, the bootstrap method results in an approximate sample from the sampling distribution of a statistic. Whereas, the composite sampling approach in Algorithm 1 can be seen as a means to sample from (\ref{bayesconverter}). This is also different from the Bayesian bootstrap \citep{rubin1981bayesian}, which does not restrict the samples to be from a posterior distribution of the form in (\ref{bayesconverter}).

\begin{algorithm}[t]\caption{Algorithm 1: Implementation of the HGT Model.}\label{euclid}
	\begin{algorithmic}[1]
		\item Set $b = 1$ and initialize $\textbf{h}$, $\bm{\gamma}$, $\textbf{y}$, and $\bm{\theta}$ with $\textbf{h}^{[0]}$, $\bm{\gamma}^{[0]}$,  $\textbf{y}^{[0]}$, and $\bm{\theta}^{[0]}$.
		\item Sample $\textbf{h}^{[b]}$ from $f(\textbf{h}\vert \bz, \bm{\gamma}^{[b-1]})$.
		\item Sample $\bm{\gamma}^{[b]}$ from their full-conditional distributions. We use the slice sampler \citep{neal2003slice} if the full-conditional does not have a closed form.
		\item Sample $\textbf{y}^{[b]}$ and $\bm{\theta}^{[b]}$ from $f(\textbf{y}, \bm{\theta}\vert \textbf{h}^{[b]})$, which is the posterior distribution associated with the preferred model described in (\ref{preferredModel}).
		\item Set $b = b+1$.
		\item Repeat Steps 2$\--$5 until $b = B$ for a prespecified value of $B$.
	\end{algorithmic}
\end{algorithm}

\subsection{Modeling the Data Given Transformations} \label{Section:3.3}
Consider the following specifications of the data models:
\begin{align}\label{transat}
\nonumber
Z_{i1} \vert h_{i1}&\sim \mathrm{Normal}(h_{i1},v)\\
\nonumber
Z_{i2} \vert h_{i2}&\sim \mathrm{Binomial}\left\lbrace b_{i},\frac{\mathrm{exp}\left(h_{i2}\right)}{1+\mathrm{exp}\left(h_{i2}\right)}\right\rbrace\\
Z_{i3} \vert h_{i3}&\sim \mathrm{Poisson}\left\lbrace\mathrm{exp}\left(h_{ij}\right)\right\rbrace; \hspace{5pt}i = 1,\ldots, I_{j}, j = 1,2,3,
\end{align}
\noindent
which is different from the GLMM in (\ref{glm}). Specifically, instead of conditioning on the latent process of interest $Y_{ij}$, we condition on the transformation $h_{ij}$.

With the transformation model $f(\textbf{z}_{trn}\vert \textbf{h})$ defined, we are left to specify a transformation prior and transformation hyperprior. We define the transformation prior to be the conjugate distributions associated with (\ref{transat}). It follows from \citet{diaconis} that the conjugate distribution for $h_{ij}$ is given by,
\begin{equation}\label{univ_LG}
f_{DY}(h_{ij}\vert \alpha_{j}, \kappa_{j},a,b) = K(\alpha_{j}, \kappa_{j})\mathrm{exp}\left\lbrace \alpha_{j} h_{ij} - \kappa_{j} \psi_{j}(h_{ij})\right\rbrace; \hspace{2pt}i = 1,\ldots, I_{j}, j = 1,\ldots, J,
\end{equation}
\noindent
where $K(\alpha_{j}, \kappa_{j})$ is a normalizing constant, $h_{ij}\in \mathbb{R}$, $\alpha_{1} \in \mathbb{R}$, $\kappa_{2}>\alpha_{2}$, $\alpha_{m}>0$, and $\kappa_{k} > 0$; for $m = 2,3,$ and $k = 1,3$. Let $\psi_{1}(Z) = Z^2$, $\psi_{2}(Z) = \mathrm{log}(1+e^{Z})$, and $\psi_{3}(Z) = \mathrm{exp}(Z)$. Also, we use the shorthand $\mathrm{DY}(\alpha_{j},\kappa_{j};\hspace{2pt}\psi_{j})$ to represent the pdf in (\ref{univ_LG}). Finally, let $\bm{\gamma} = (\alpha_{1},\alpha_{2},\alpha_{3}, \kappa_{1},\kappa_{2},\kappa_{3}, v)^{\prime}$ be the transformation hyperparameter. The DY distribution is a special case of the recently introduced conjugate multivariate distribution \citep{bradleyLCM}, where the matrix-valued covariance parameter is set equal to the identity matrix.

Equations (\ref{transat}) and (\ref{univ_LG}) can be used to produce a full-conditional distribution for the elements of $\textbf{h}$:
\begin{align}\label{saturatedModel}
\nonumber
h_{i1}\vert Z_{i1}, \bm{\gamma}  &\sim \mathrm{Normal}\left\lbrace \left(2\kappa_{1}+\frac{1}{v}\right)^{-1}\left(\frac{Z_{i1}}{v} + \alpha_{1}\right), \left(2\kappa_{1}+\frac{1}{v}\right)^{-1}\right\rbrace; \hspace{5pt}i = 1,\ldots, I_{1}\\
\nonumber
h_{i2}\vert Z_{i2}, \bm{\gamma}  &\sim \mathrm{DY}\left( \alpha_{2} + Z_{i2}, \kappa_{2} + b_{i};\hspace{2pt} \psi_{2}\right); \hspace{5pt}i = 1,\ldots, I_{2}\\
h_{i3}\vert Z_{i3}, \bm{\gamma}  &\sim \mathrm{DY}\left( \alpha_{3} + Z_{i3}, \kappa_{3} + 1;\hspace{2pt} \psi_{3}\right); \hspace{5pt}i = 1,\ldots, I_{3}.
\end{align}
\noindent
The derivations of the full conditional distributions are fairly straightforward, and are given in Appendix A. One can simulate directly from the posterior distribution in (\ref{saturatedModel}). Replicates of $h_{ij}$ from (\ref{saturatedModel}) can be computed using the following transformation \citep{bradleyLCM}:
\begin{align}\label{Step2}
\nonumber
h_{i1} &\overset{d}{=} \left(2\kappa_{1}+\frac{1}{v}\right)^{-1}\left(\frac{Z_{i1}}{v} + \alpha_{1}\right) + w_{1}; \hspace{5pt}i = 1,\ldots, I_{1}\\
\nonumber
h_{i2} &\overset{d}{=}  \mathrm{log}\left(\frac{w_{2}}{1-w_{2}}\right); \hspace{5pt}i = 1,\ldots, I_{2}\\
h_{i3} &\overset{d}{=} \mathrm{log}\left(w_{3}\right); \hspace{5pt}i = 1,\ldots, I_{3},
\end{align}
where ``$\overset{d}{=}$'' stands for equal in distribution, $w_{1}\vert Z_{i1},\alpha_{1},\kappa_{1},v$ is distributed normally with mean zero and variance $\left(2\kappa_{1}+\frac{1}{v}\right)^{-1}$, $w_{2}\vert Z_{i2},\alpha_{2},\kappa_{2}$ is distributed according to a beta distribution with shape parameters $(\alpha_{2}+Z_{i2})$ and $(\kappa_{2}-\alpha_{2}+b_{i}-Z_{i2})$, and $w_{3}\vert Z_{i3},\alpha_{3},\kappa_{3}$ is distributed according to a gamma distribution with shape parameter $(\alpha_{3}+Z_{i3})$ and rate parameter $(\kappa_{3}+1)$. Step 2 of Algorithm 1 involves simulating according to (\ref{Step2}), which is straightforward.

The specification of a transformation hyperprior for $\bm{\gamma}$ is crucial to guarantee that $f(h_{ij}\vert Z_{ij},\bm{\gamma})$ is proper in the event that $Z_{i3} = 0$, $Z_{i2} = 0$, or $Z_{i3} = b_{i}$. Thus, we assume $\alpha_{1}=\kappa_{1} = 0$, $\alpha_{2}$ and $\alpha_{3}$ are distributed according to a gamma distribution, $\kappa_{2}\vert \alpha_{2}$ is distributed according to a shifted (by $\alpha_{2}$) gamma distribution, $\kappa_{3}$ follows a gamma distribution, and $v$ is distributed according to an inverse gamma distribution \citep[e.g., see][among others]{gelmanprior}. These transformation hyperpriors are explicitly stated, and the full-conditional distributions for $\bm{\gamma}$ are derived in Appendix B.1.
%

Section~\ref{Section:3.2} is flexible enough to allow for a transformation prior that implies cross-dependence among the elements of $\textbf{h}$, but we do not consider this case in this article. The main reason for this choice is that transformations are used in place of the original data set when implementing the preferred model (Step 4 of Algorithm 1). That is, the transformed values are used as a proxy for (or in place of) the data in the preferred model. Consequently, we would like to specify $\textbf{h}$ to ``overfit'' the data so that $\textbf{h}$ can reasonably be thought of as a proxy for the data.

 Our choice of the prior in (\ref{univ_LG}) leads to posterior replicates that overfit the data. In particular, it is straightforward to verify that 
\begin{align}\label{overfitit}
\nonumber
&\underset{\kappa_{1}\rightarrow 0}{\mathrm{lim}}\hspace{2pt}\underset{\alpha_{1}\rightarrow 0}{\mathrm{lim}}E\left\lbrace h_{i1}\vert Z_{i1}, \bm{\gamma}\right\rbrace = Z_{i1}\\ &\underset{\kappa_{2}\rightarrow 0}{\mathrm{lim}}\hspace{2pt}\underset{\alpha_{2}\rightarrow 0}{\mathrm{lim}}E\left\lbrace b_{j}g_{2}^{-1}(h_{j2})\vert Z_{j2}, \bm{\gamma}\right\rbrace = Z_{j2}\\
\nonumber
&\underset{\kappa_{3}\rightarrow 0}{\mathrm{lim}}\hspace{2pt}\underset{\alpha_{3}\rightarrow 0}{\mathrm{lim}}E\left\lbrace g_{3}^{-1}(h_{k3})\vert Z_{k3}, \bm{\gamma}\right\rbrace = Z_{i3};\hspace{5pt}i = 1,\ldots, I_{1}, j = 1,\ldots, I_{2}, k = 1,\ldots, I_{3}.
\end{align}
\noindent
See Appendix A for the derivation of (\ref{overfitit}). Thus, the posterior mean of $\textbf{h}$ (on the original scale of the data) are exactly the observed data $\{Z_{ij}\}$ as the hyperparameters go to zero. This suggests that estimates from $f(\textbf{h}\vert \textbf{z}_{trn})$ overfits the data, however, it is not necessarily true that $f(\textbf{y},\bm{\theta}\vert \textbf{z}_{trn})$ overfits the data.

\subsection{Example of Bayesian Implementation}\label{Section:3.4}
 Consider the following mixed effects model for the transformed data \citep[e.g., see][among others]{johan}:
\begin{align}\label{summary33}
\nonumber
&\mathrm{Transformed\hspace{5pt}Data\hspace{5pt}Model:}\hspace{5pt}h_{ij}\vert \bfbeta, \bm{\eta}, \xi_{ij},\bm{\lambda}\ind \mathrm{Normal}\left(\textbf{x}_{ij}^{\prime}\bfbeta + \textbf{S}_{ij}^{\prime}\bm{\eta} + \xi_{ij}, \sigma^{2}\right)\hspace{5pt}m(\textbf{h}\vert \bm{\lambda});\\
\nonumber
\nonumber
&\mathrm{Process\hspace{5pt}Model\hspace{5pt}1:}\hspace{5pt} \bm{\eta}\vert \sigma_{\eta}^{2} \sim \mathrm{Normal}\left(\bm{0}_{r}, \sigma_{\eta}^{2}\textbf{I}_{r}\right);\\
\nonumber
&\mathrm{Process\hspace{5pt}Model\hspace{5pt}2:}\hspace{5pt} \xi_{ij}\vert \sigma_{\xi}^{2} \ind \mathrm{Normal}\left(0, \sigma_{\xi}^{2}\right);\\
\nonumber
&\mathrm{Prior\hspace{5pt}1:}\hspace{5pt} \sigma^{2} \sim \mathrm{IG}\left(\alpha_{v}, \beta_{v}\right);\\
\nonumber
&\mathrm{Prior\hspace{5pt}2:}\hspace{5pt} \bfbeta \sim \mathrm{Normal}\left(\bm{0}_{p}, \sigma_{\beta}^{2}\textbf{I}_{p}\right);\\
\nonumber
&\mathrm{Prior\hspace{5pt}3:}\hspace{5pt} \sigma_{\xi}^{2} \sim \mathrm{IG}\left(\alpha_{\xi}, \beta_{\xi}\right);\\
&\mathrm{Prior\hspace{5pt}4:}\hspace{5pt} \sigma_{\eta}^{2} \sim \mathrm{IG}\left(\alpha_{\eta}, \beta_{\eta}\right);\hspace{2pt}i = 1,\ldots I_{j}, j = 1,2,3,
\end{align}
\noindent
where $\textbf{x}_{ij}$ is a $p$-dimensional vector of known covariates, $\textbf{I}_{r}$ is a $r\times r$ identity matrix, $\bm{0}_{r}$ is an $r$-dimensional vector of zeros, $\alpha_{v}=\alpha_{\eta}=\alpha_{\xi}=1$, $\beta_{v}=\beta_{\eta}=\beta_{\xi}=1$, $\sigma_{\beta}^{2} = 100$, and $\bm{\xi}=(\xi_{11},\ldots, \xi_{I_{3}3})^{\prime}$. The hyperparameters are chosen so that the prior is relatively ``flat'' and we find that our results are robust to these specifications. In Algorithm 1, we set $Y_{ij} = \textbf{x}_{ij}^{\prime}\bm{\beta}+\textbf{S}_{ij}^{\prime}\bm{\eta}+\xi_{ij}$ and $\bm{\theta} = (\bm{\beta}^{\prime}, \sigma^{2}, \sigma_{\xi}^{2}, \sigma_{\eta}^{2})^{\prime}$. The choice of basis functions and specification of $r$ are important. In Appendix B.2, we give these details. 

The full conditional distributions for $\textbf{y}$ and $\bm{\theta}$ are well-known \citep[e.g.,see][for a standard reference reference]{cressie-wikle-book} and are listed in Appendix B.3. Thus, Step 2 of Algorithm 1 involves simulating according to (\ref{Step2}) and Step 4 of Algorithm 1 involves sequentially simulating simulating from these standard full-conditional distributions. Details are given in Appendix B.1 and B.3.

\section{Statistical Inference}\label{Section:4}

 Estimation and prediction over the training set can be done by computing summary statistics using the quantities generated in Step 4 of Algorithm 1. However, to forecast values (e.g., future cases or deaths due to COVID-19) we make use of validation and testing datasets, which is a common strategy in machine learning \citep{htf}.

\subsection{Estimation and Goodness-of-Fit using Training Data} \label{Section:4.1}

Estimation and prediction of $Y_{ij}$ at $i = 1,\ldots, I_{j}$ is rather natural using the output from Algorithm 1. In particular, let $\bm{\theta}^{[b]}$ and $Y_{ij}^{[b]}$ be the $b$-th replicate from Step 4 in Algorithm 1. Then one can estimate $\bm{\theta}$ and $Y_{ij}$ using summary statistics such as:
\begin{align*}
\widehat{E}(Y_{ij}\vert \textbf{z}_{trn}) &= \frac{1}{B-b_{0}}\sum_{b=b_{0} +1}^{B}\sum_{b = b_{0}}^{B} Y_{ij}^{[b]};\hspace{5pt} i = 1,\ldots; I_{j}, j = 1,2,3\\
\widehat{E}(\bm{\theta}\vert \textbf{z}_{trn}) &= \frac{1}{B-b_{0}}\sum_{b=b_{0} +1}^{B}\sum_{b = b_{0}}^{B} \bm{\theta}^{[b]},
\end{align*}
among several other summary statistics are also computed in our analyses (in our analyses we also compute percentiles to asses uncertainty). Here, $b_{0}$ is a ``burn-in'' value. In the context of the linear model in Section~\ref{Section:3.4}, we would be interested in summary statistics of $\sum_{i: t_{i} = t}\sum_{j} \textbf{S}_{ij}^{\prime}\bm{\eta}$, where recall $\bm{\eta}$ is the random effect that is shared across response types. Estimates of this random effect can be used to summarize the relationship between response types.

Assessment of the goodness of fit can be done similar to residual analyses of transformed data in traditional regression analyses. We compute the residuals $\bm{\delta} = \left(\delta_{ij}:i = 1,\ldots, I_{j}, j = 1,2,3\right)^{\prime}$, $\delta_{ij} = h_{ij}-Y_{ij}$, and compute a credible region associated with $\bm{\delta}$ \citep[e.g., see][for a standard reference]{gelmanbook}. For example, for each $i$ and $j$, find the values $q_{L,ij}$ and $q_{U,ij}$, where
\begin{equation}\label{credible_intervals}
\int_{q_{L,ij}}^{q_{U,ij}}f(\delta_{ij}\vert \bz) d\delta_{ij} = 1-\alpha; i = 1,\ldots, I_{j}, j = 1,\ldots, J,
\end{equation}
\noindent
and where $\alpha$ is prespecified. A default choice is $\alpha = 0.05$. In practice, it is rather straightforward to approximate $q_{L,ij}$ and $q_{U,ij}$. Let $h_{ij}^{[b]}$ and $Y_{ij}^{[b]}$ be the $b$-th posterior replicate of $h_{ij}$ and $Y_{ij}$ so that $\delta_{ij}^{[b]} = h_{ij}^{[b]} -Y_{ij}^{[b]}$ is the $b$-th posterior replicate of $\delta_{ij}$. Then $q_{L,ij}$ and $q_{U,ij}$ can be approximated with the $\alpha/2$ and $1-\alpha/2$ percentiles of the set $\{\delta_{ij}^{[b]}: b = 1,\ldots, B\}$, respectively. If the value of zero lies within this interval (e.g., $q_{L,ij}<0<q_{U,ij}$) for many values of $i$ and $j$, then this suggests that the model for $\by$ provides a reasonable fit to this data set.

Equation (\ref{overfitit}) shows that the posterior mean of the transformation models overfits the data, which we motivated as a way to avoid oversmoothing estimates of $\textbf{y}$ and $\bm{\theta}$ in Algorithm 1. However, the fact that the transformation model overfits is also important from the point-of-view of diagnostics. In particular, in the goodness-of-fit literature, overfitted values are often used as a proxy for the data. For example, in log-linear models the most parsimonious reduced model that is not significantly different (in terms of the deviance or chi-square statistic) from the saturated model (an overfitted model) is used for statistical inference \citep[e.g., see][for a standard reference]{agresti2011categorical}. This is exciting because it provides a new way to conduct classical residual analysis in a Bayesian multi-response context. In particular, in Sections~\ref{Section:5} we give an example of plotting the (posterior median) residuals versus a useful covariate not included in the analysis to assess whether or not it should be included in a model.

\subsection{Estimating Hyperparamers using a Validation Dataset} In machine learning, one often adjusts the model for being biased towards the training data by holding aside a dataset to estimate hyperparameters. This hold-out dataset is referred to as a validation dataset \citep{htf}. The validation dataset $\textbf{z}_{val} = (Z_{ij}: i = I_{j}^{val}+1, \ldots, I, j = 1,2,3)^{\prime}$ is observed over the indices $i \in \{I_{j}+1,\ldots, I_{j}^{val}\}$ and $j =1,2,3$, where $I_{j}<I_{j}^{val}\le I$. Additionally, let $Y_{ij}^{*}$ be different from, but independent and identically distributed as $Y_{ij}$. We can not replace $Y_{ij}^{*}$ with $Y_{ij}$ in our analysis of the validation data, otherwise, the validation data would be included with the training data when updating $Y_{ij}$. Then, we assume
\begin{align}\label{glm_cross}
\nonumber
Z_{i1} \vert Y_{i1}^{*} &\sim \mathrm{Normal}(k_{1}(Y_{i1}^{*}, \bm{\kappa}),v)\\
\nonumber
Z_{i2} \vert Y_{i2}^{*} &\sim \mathrm{Binomial}\left\lbrace b_{i},k_{2}(Y_{i2}^{*}, \bm{\kappa})\right\rbrace\\
Z_{i3} \vert Y_{i3}^{*} &\sim \mathrm{Poisson}\left\lbrace k_{3}(Y_{i3}^{*}, \bm{\kappa})\right\rbrace; \hspace{5pt}i = I_{j}+1,\ldots, I_{j}^{val}\\
f(\bm{\kappa})&,
\end{align}
where $\bm{\kappa}$ is a generic $d$-dimensional vector of real-valued parameters and $f(\bm{\kappa})$ is the prior distribution of this parameter. The functions $k_{j}(Y_{ij},\bm{\kappa})$ are not necessarily equal to $g_{j}(Y_{ij})$, and we parameterize the unknown function $k_{j}$ with $\bm{\kappa}$. In this article, we allow for either $k_{j} = g_{j}$ so that $\bm{\kappa} \equiv 0$, or $g_{j}$ to be adjusted linearly so that,
\begin{equation}
k_{j}(Y) = \kappa_{j0} + \kappa_{j1}g_{j}(Y);\hspace{5pt} j = 1,2,3, \hspace{2pt}Y\in \mathbb{R},
\end{equation}
\noindent
and $\bm{\kappa} = (\kappa_{10}, \kappa_{20}, \kappa_{30},\kappa_{11}, \kappa_{21}, \kappa_{31})^{\prime}$.  In this setting, we choose the improper flat prior $f(\bm{\kappa}) = 1$. When $k_{j} = g_{j}$ so that $\bm{\kappa} \equiv 0$ there is no need to consider a validation dataset, since there is no hyperparameter $\bm{\kappa}$ to estimate.
\begin{algorithm}[t]\caption{Algorithm 2: Steps Needed for Fitting the Validation Data.}\label{euclid2}
	\begin{algorithmic}[1]
		\item Set $b = 1$ and initialize $Y_{ij}^{*}$ and $\bm{\kappa}$ with $Y_{ij}^{*[0]}$ and $\bm{\kappa}^{[0]}$.
		\item Sample $Y_{ij}^{*[b]}$ using Algorithm 1
		\item Sample $\bm{\kappa}^{[b]}$ from it's full-conditional distribution. We use the slice sampler \citep{neal2003slice} since the full-conditional distribution does not have a closed form.
		\item Set $b = b+1$.
		\item Repeat Steps 2$\--$5 until $b = B$ for a prespecified value of $B$.
	\end{algorithmic}
\end{algorithm}

\begin{algorithm}[t]\caption{Algorithm 3: Steps Needed for Forecasting.}\label{euclid3}
	\begin{algorithmic}[1]
		\item Set $b = 1$ and initialize $Y_{ij}^{**}$ and $\bm{\kappa}^{*}$ with $Y_{ij}^{**[0]}$ and $\bm{\kappa}^{*[0]}$.
		\item Sample $Y_{ij}^{**[b]}$ using Algorithm 1.
		\item Sample $\bm{\kappa}^{*[b]}$ using Algorithm 2.
		\item Sample $Z_{ij}^{[b]}$ from (\ref{glmPP}).
		\item Set $b = b+1$.
		\item Repeat Steps 2$\--$5 until $b = B$ for a prespecified value of $B$.
		\item Compute the sample mean and variance (across the index $b$) of $Z_{ij}^{[b]}$.
	\end{algorithmic}
\end{algorithm}
\subsection{Forecasting} We produce next day forecasts for the variables in our study. In particular, the testing observations are defined over the indices $i = I_{j}^{val}+1,\ldots, I$ for $j = 1,2,3$. We let $\bm{\kappa}^{*}$ and $Y_{ij}^{**}$ be distributed according to $f(\bm{\kappa}\vert \textbf{z}_{trn}, \textbf{z}_{val})$ and $f(Y_{ij}\vert \textbf{z}_{trn})$, respectively. Again, we can not let $Y_{ij}^{**}$ equal $Y_{ij}$, since otherwise, the testing data would be included when updating $Y_{ij}$ based on the training data. Then, we assume that
\begin{align}\label{glmPP}
\nonumber
Z_{i1} \vert Y_{i1}^{**}, \bm{\kappa}^{*}&\sim \mathrm{Normal}(k_{1}(Y_{i1}^{**},\bm{\kappa}^{*}),v)\\
\nonumber
Z_{i2}\vert Y_{i2}^{**}, \bm{\kappa}^{*} &\sim \mathrm{Binomial}\left\lbrace b_{i},k_{2}(Y_{i2}^{**},\bm{\kappa}^{*})\right\rbrace\\
Z_{i3} \vert Y_{i3}^{**}, \bm{\kappa}^{*} &\sim \mathrm{Poisson}\left\lbrace k_{1}(Y_{i3}^{**},\bm{\kappa}^{*})\right\rbrace; \hspace{5pt}i = I_{j}^{val},\ldots, I, j = 1,2,3.
\end{align}
Predictions of the data process and estimation of cross-covariances can be found using in a similar manner as (\ref{predZ}) and (\ref{covZZ}). That is, the posterior mean and covariance of $Z_{ij}$ and $Z_{km}$ is, $E(Z_{ij}\vert \textbf{z}_{trn})$ and $cov(Z_{ij},Z_{km}\vert \textbf{z}_{trn})$, where recall, under the mixed effects assumption $cov(Z_{ij},Z_{km}\vert \textbf{z}_{trn}) = \textbf{S}_{ij}^{\prime}cov(\bm{\eta}\vert \textbf{z}_{trn})\textbf{S}_{ij}$, which is not necessarily zero. Implementation is be summarized in Algorithm 3. When $g_{j} \equiv k_{j}$ and $\bm{\kappa}\equiv 0$, the predictions and covariances are simply  
\begin{align}
\label{predZpred}
E(Z_{ij}\vert \textbf{z}_{trn}) &= E\left\lbrace c_{ij}g_{j}^{-1}(Y_{ij})\vert \textbf{z}_{trn}\right\rbrace\\
\nonumber
cov(Z_{ij},Z_{mk}\vert \textbf{z}_{trn}) &= cov(c_{ij}g_{j}^{-1}(Y_{ij}),c_{mk}g_{j}^{-1}(Y_{mk})\vert \textbf{z}_{trn}),
\end{align}
\noindent
which can be directly computed from Step 4 of Algorithm 1. Once the next day data is observed, it is treated as ``testing data,'' which is then used to assess the performance of our forecasts (e.g., through the root mean squared error, etc.).

\subsection{Summaries of the Models used for Inference}
There are three models used to do statistical inference, one that uses the training data, another based on validation data, and a third based on testing data. The joint distribution of the training data, processes, and parameters is written as the product of the following conditional distributions:
	\begin{align}
	\nonumber
	&\mathrm{Training\hspace{5pt}Data\hspace{5pt}Model\hspace{5pt}1:}\hspace{5pt} Z_{i1} \vert h_{i1}\sim \mathrm{Normal}(h_{i1},v)\\
	\nonumber
	&\mathrm{Training\hspace{5pt}Data\hspace{5pt}Model\hspace{5pt}2:}\hspace{5pt}Z_{i2} \vert h_{i2}\sim \mathrm{Binomial}\left\lbrace b_{i},\frac{\mathrm{exp}\left(h_{i2}\right)}{1+\mathrm{exp}\left(h_{i2}\right)}\right\rbrace\\
	\nonumber
	&\mathrm{Training\hspace{5pt}Data\hspace{5pt}Model\hspace{5pt}3:}\hspace{5pt}Z_{i3} \vert h_{i3}\sim \mathrm{Poisson}\left\lbrace\mathrm{exp}\left(h_{ij}\right)\right\rbrace; \hspace{5pt}i = 1,\ldots, I_{j}, j = 1,2,3\\
	\label{summary4}
	&\mathrm{Transformed\hspace{5pt}Data\hspace{5pt}Model:}\hspace{5pt} f(h_{ij}\vert Y_{ij},\bm{\theta})m(\textbf{h}\vert \bm{\gamma}); \hspace{5pt}i = 1,\ldots, I_{j}, j = 1,2,3\\
	\nonumber
	&\mathrm{Process\hspace{5pt}Model:}\hspace{5pt} f(\textbf{y}\vert \bm{\theta})\\
	\nonumber
&\mathrm{Prior:}\hspace{5pt} f(\bm{\theta})\\
	\nonumber
	&\mathrm{Transformation\hspace{5pt}Hyperprior:}\hspace{5pt} f(\bm{\gamma}).
	\end{align}
\noindent
The model in (\ref{summary4}) is the aforementioned HGT model. This is a well defined proper model (see Appendix A for these details), provided that $f(h_{ij}\vert \bm{\theta})$, $f(\textbf{y}\vert \bm{\theta})$, and $f(\bm{\theta})$ are proper. 

Recall that one motivation for the model in (\ref{summary4}) is that one can incorporate their preferred model for continuous data directly into our framework. This is especially important to aid researchers in analyzing COVID-19 using their preferred approach (cite) in a computationally efficient manner, since Algorithm 1 does not require one to change the implementation of their preferred model. This flexibility arises in the data scientist's specification of $f(h_{ij}\vert \bm{\theta})$, $f(\textbf{y}\vert \bm{\theta})$, and $f(\bm{\theta})$. In Section~\ref{Section:3.3} we specify $f(h_{ij}\vert \bm{\theta})$, $f(\textbf{y}\vert \bm{\theta})$, and $f(\bm{\theta})$ using a mixed effects model, and in Section~\ref{Section:5} we also consider using BART to illustrate this flexibility. Although we only consider Bayesian specifications of the preferred model, Step 4 can easily be substituted with replicates/estimates of $\textbf{y}$ and $\bm{\theta}$ (computed using $\textbf{h}^{[b]}$) from empirical Bayesian models, approximate Bayesian models, or frequentist models.

The LCM is explicitly used in the HGT model in (\ref{summary4}) through the term $m(\textbf{h}\vert\textbf{y})$, where recall
\begin{equation*}
m(\textbf{h}\vert\textbf{y})=\frac{\prod_{i,j}f_{DY}(h_{ij}\vert \alpha_{j}, \kappa_{j}, a, b)}{\int\int(f(\textbf{h}\vert \textbf{y}, \bm{\theta})f(\textbf{y}\vert \bm{\theta})f(\bm{\theta})d\textbf{y}d\bm{\theta}},
\end{equation*}
\noindent
$\bm{\gamma} = (\alpha_{1},\alpha_{2}, \alpha_{3}, \kappa_{1}, \kappa_{2}, \kappa_{3}, a,b)^{\prime}$ , and the prior for $\bm{\gamma}$ is defined in Appendix B.1. Recall that Algorithm 1 is a collapsed Gibbs sampler, where we update $\textbf{h}$ and $\bm{\gamma}$ using the marginal distribution of (\ref{summary4}) found by integrating our $\textbf{y}$ and $\bm{\theta}$. Specifically, when integrating our $\textbf{y}$ and $\bm{\theta}$ in (\ref{summary4}), we obtain
\begin{align}
	\nonumber
	&\mathrm{Training\hspace{5pt}Data\hspace{5pt}Model\hspace{5pt}1:}\hspace{5pt} Z_{i1} \vert h_{i1}\sim \mathrm{Normal}(h_{i1},v)\\
	\nonumber
	&\mathrm{Training\hspace{5pt}Data\hspace{5pt}Model\hspace{5pt}2:}\hspace{5pt}Z_{i2} \vert h_{i2}\sim \mathrm{Binomial}\left\lbrace b_{i},\frac{\mathrm{exp}\left(h_{i2}\right)}{1+\mathrm{exp}\left(h_{i2}\right)}\right\rbrace\\
	\nonumber
	&\mathrm{Training\hspace{5pt}Data\hspace{5pt}Model\hspace{5pt}3:}\hspace{5pt}Z_{i3} \vert h_{i3}\sim \mathrm{Poisson}\left\lbrace\mathrm{exp}\left(h_{ij}\right)\right\rbrace; \hspace{5pt}i = 1,\ldots, I_{j}, j = 1,2,3\\
		\nonumber
	&\mathrm{Transformation\hspace{5pt}Prior:}\hspace{5pt} \prod_{i,j}f_{DY}(h_{ij}\vert \alpha_{j}, \kappa_{j}, a, b)\\
	\nonumber
	&\mathrm{Transformation\hspace{5pt}Hyperprior:}\hspace{5pt} f(\bm{\gamma}).
\end{align}
\noindent
which leads to the computationally simple updates of $\textbf{h}$ and $\bm{\theta}$ developed in Section~\ref{Section:3.3} to be used in Step 2 of Algorithm 1.

The joint distribution of the validation data, processes, and parameters is written as the product of the following conditional distributions:
\begin{subequations}
	\begin{align}
	\label{summary1}
	&\mathrm{Validation\hspace{5pt}Data\hspace{5pt}Model\hspace{5pt}1:}\hspace{5pt}Z_{i1} \vert Y_{i1}^{*},\bm{\kappa} \sim \mathrm{Normal}(k_{1}(Y_{i1}^{*}, \bm{\kappa}),v)\\
	\label{summary2}
	&\mathrm{Validation\hspace{5pt}Data\hspace{5pt}Model\hspace{5pt}2:}\hspace{5pt}Z_{i2} \vert Y_{i2}^{*},\bm{\kappa} \sim \mathrm{Binomial}\left\lbrace b_{i},k_{2}(Y_{i2}^{*}, \bm{\kappa})\right\rbrace\\
		\label{summary3}
	&\mathrm{Validation\hspace{5pt}Data\hspace{5pt}Model\hspace{5pt}3:}\hspace{5pt}Z_{i3} \vert Y_{i3}^{*},\bm{\kappa} \sim \mathrm{Poisson}\left\lbrace k_{3}(Y_{i3}^{*}, \bm{\kappa})\right\rbrace\\
\nonumber
	&\mathrm{Posterior \hspace{5pt}Process\hspace{5pt}Model:}\hspace{5pt} f(Y_{ij}^{*}\vert \textbf{z}_{trn})\\
	\nonumber
	&\mathrm{Prior:}\hspace{5pt} f(\bm{\kappa}); \hspace{5pt}i = I_{j}+1,\ldots, I_{j}^{val}, j = 1,2,3,
	\end{align}
\end{subequations}
\noindent
where recall that the goal of this model is to estimate $\bm{\kappa}$ from its posterior $f(\bm{\kappa}\vert \textbf{z}_{val}, \textbf{z}_{trn})$, which is a parameter that allows one to avoid overfitting the training data. The distribution $f(Y_{ij}^{*}\vert \textbf{z}_{trn})$ is the posterior distribution implied by the model in (\ref{summary4}). Model (24) can be implemented through Algorithm 2. When $f(\textbf{y}\vert \bm{\theta})$ is specified according to a linear model (i.e., Equation (\ref{glmm})) then Equations (\ref{summary1}) through (\ref{summary3}) can be thought of as a GLMM \citep{glm-mcculloch}. GLMMs also arise in our model for testing data. The joint distribution of the testing data, processes, and parameters is written as the product of the following conditional distributions:
	\begin{align}
	\nonumber
	&\mathrm{Testing\hspace{5pt}Data\hspace{5pt}Model\hspace{5pt}1:}\hspace{5pt} Z_{i1} \vert Y_{i1}^{**},\bm{\kappa}^{*} \sim \mathrm{Normal}(k_{1}(Y_{i1}^{**}, \bm{\kappa}^{*}),v)\\
	\nonumber
	&\mathrm{Testing\hspace{5pt}Data\hspace{5pt}Model\hspace{5pt}2:}\hspace{5pt}Z_{i2} \vert Y_{i2}^{**},\bm{\kappa}^{*} \sim \mathrm{Binomial}\left\lbrace b_{i},k_{2}(Y_{i2}^{**}, \bm{\kappa}^{*})\right\rbrace\\
	\label{summary31}
	&\mathrm{Testing\hspace{5pt}Data\hspace{5pt}Model\hspace{5pt}3:}\hspace{5pt}Z_{i3} \vert Y_{i3}^{**},\bm{\kappa}^{*} \sim \mathrm{Poisson}\left\lbrace k_{3}(Y_{i3}^{**}, \bm{\kappa}^{*})\right\rbrace\\
	\nonumber
	&\mathrm{Posterior \hspace{5pt}Process\hspace{5pt}Model:}\hspace{5pt} f(Y_{ij}^{**}\vert \textbf{z}_{trn})\\
	\nonumber
	&\mathrm{Posterior \hspace{5pt}Parameter\hspace{5pt} Model:}\hspace{5pt} f(\bm{\kappa}^{*}\vert \textbf{z}_{val},\textbf{z}_{trn}); \hspace{5pt}i = I_{j}^{val}+1,\ldots, I, j = 1,2,3,
	\end{align}
\noindent
where the goal is to predict $Z_{ij}$ at $i = I_{j}^{val}+1,\ldots, I$ and $j = 1,2,3$. The distribution $f(Y_{ij}^{**}\vert \textbf{z}_{trn})$ is the posterior distribution implied by the model in (\ref{summary4}) and $f(\bm{\kappa}^{*}\vert \textbf{z}_{val},\textbf{z}_{trn})$ is the posterior distribution from (24). Model (\ref{summary31}) can be implemented through Algorithm 3. For example, $Z_{ij}$ in Section~\ref{Section:6} is the number of observed cases, deaths, and recoveries from COVID-19 in April 8, 2020, and the posterior predictions from the model in (\ref{summary31}) represent the next day forecasts.

\section{Simulations} \label{Section:5}

The goals of this simulation study is to provide a standard demonstration that the HGT model produces reasonable predictions. Another goal is to illustrate the flexibility of the HGT model to specify a data scientist's preferred model for continuous data. To do this we apply (\ref{summary4}) to the spatio-temporal mixed effects model in Section~\ref{Section:3.4} and BART (details in Appendix B.4).

\subsection{Simulation Setup}\label{Section:5.1}

\citet{friedman1991multivariate} introduced a simulation design, which has become a useful benchmark study \citep[e.g., see][among others]{chipman2010bart}. Let
\begin{equation}\label{Fried}
h(x_{1,ij},\ldots, x_{10,ij}) = 10 \mathrm{sin}(\pi x_{1,ij} x_{2,ij}) + 20 (x_{3,i}- 0.5)^{2} + 10x_{4,ij} + 5x_{5,i}; i = 1,\ldots, I, j = 1,2,3,
\end{equation}
\noindent
which includes two non-linear terms, two linear terms, and a non-linear interaction. We consider the following specifications of the distributional assumptions associated with the data:
\begin{align}\label{simdata}
Z_{i1} &\sim \mathrm{Normal}(h(x_{1,i1},\ldots, x_{10,i1}),1)\\
\nonumber
Z_{i2} &\sim \mathrm{Binomial}\left\lbrace 300,\frac{\mathrm{exp}\left(h(x_{1,i2},\ldots, x_{10,i2})\right)}{1+\mathrm{exp}\left(h(x_{1,i2},\ldots, x_{10,i2})\right)}\right\rbrace\\
\nonumber
Z_{i3} &\sim \mathrm{Poisson}\left\lbrace\mathrm{exp}\left(h(x_{1,i3},\ldots, x_{10,i3})\right)\right\rbrace,
\end{align}
\noindent
for $i = 1,\ldots, I_{j}$. Methods are compared using the root mean squared error (RMSE),
\begin{equation*}
\left(\frac{\sum_{i = 1}^{I}\sum_{j = 1}^{3}\left[\widehat{g}_{j}^{\hspace{1pt}-1}\left\lbrace h(x_{1,ij},\ldots, x_{10,ij})\right\rbrace - g_{j}^{-1}\left\lbrace h(x_{1,ij},\ldots, x_{10,ij}) \right\rbrace \right]^{2}}{3I}\right)^{1/2},
\end{equation*}
\noindent
where $\widehat{g}_{j}^{\hspace{1pt}-1}(h)$ is estimated using Monte-Carlo integration using 2,000 iterations with a burn-in of 1,000. For each Bayesian method, we let $\widehat{g}_{j}^{\hspace{1pt}-1}(h)$ be the pointwise posterior mean of ${g}_{j}^{\hspace{1pt}-1}(h)$. We fit the preferred model using covariates $x_{1,ij},x_{3,ij},x_{4,ij},\ldots, x_{10,ij}$, and hence, we consider the case were an important covariate is not observed (i.e., $\{x_{2,ij}\}$) and several unneeded covariates are included (i.e., $\{x_{6,ij},\ldots, x_{10,ij}\}$ are not present in (\ref{Fried})). The omissions of $\{x_{2,ij}\}$ when implementing our method is a slight departure from the original setup in \citet{friedman1991multivariate}. However, we feel that it is more realistic to assume that not all covariates are observed in practice, and will be a helpful choice for illustration. We specify $x_{k,ij}\sim \mathrm{Uniform}(0,1)$, where $\mathrm{Uniform}(0,1)$ is a shorthand for the uniform distribution over the interval $[0,1]$ and $k = 1,\ldots, 10$. The preferred models are spatio-temporal mixed effects and BART (and an extension), whose implementation are described in Appendix B.3 and Appendix B.4, respectively. Additionally, the choice of basis functions are described in Appendix B.1. In the implementation of each preferred method, we allow each response type to have different regression coefficients.

\begin{figure}[t]
	\begin{center}
		\begin{tabular}{c}
			\includegraphics[width=10cm,height=8cm]{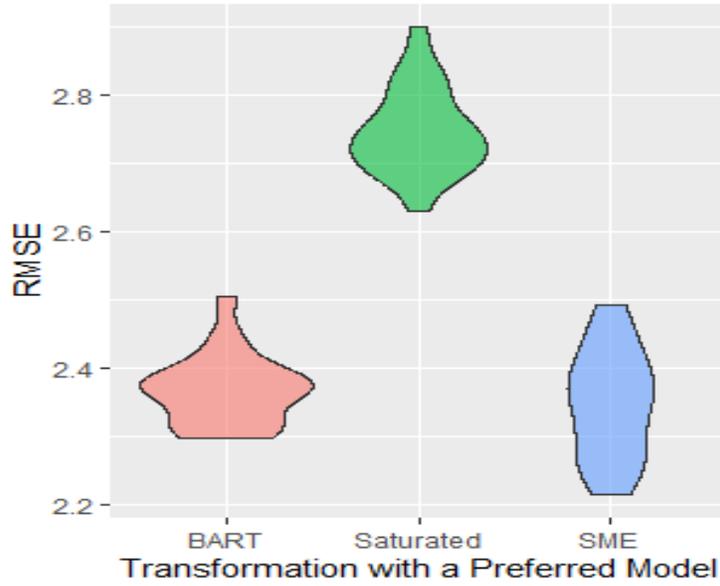}
		\end{tabular}
		\caption{A violin plot of the RMSE (y-axis) by method (x-axis) over 20 independent replicates of the data. The data are simulated as described in Section~\ref{Section:5.1}. Each method is implemented using Algorithm 1, except the method ``Saturated.'' }\label{fig:1}
	\end{center}
\end{figure}
\subsection{Simulations: Joint Analysis of Multiple Response Types}In this section, we evaluate the predictive performance of our Bayesian model with unknown transformations in the multi-response setting. In particular, we set the preferred model equal to BART \citep{chipman2010bart} and a Bayesian version of the spatio-temporal mixed effects model \citep{johan} using basis functions introduced by \citep{hughes}. The posterior mean of $h_{ij}$ (referred to as the saturated model) are included as a default poor estimator, since it is known to overfit the data (see Proposition 3).

\begin{figure}[t]
	\begin{center}
		\begin{tabular}{cc}
			\includegraphics[width=8cm,height=8cm]{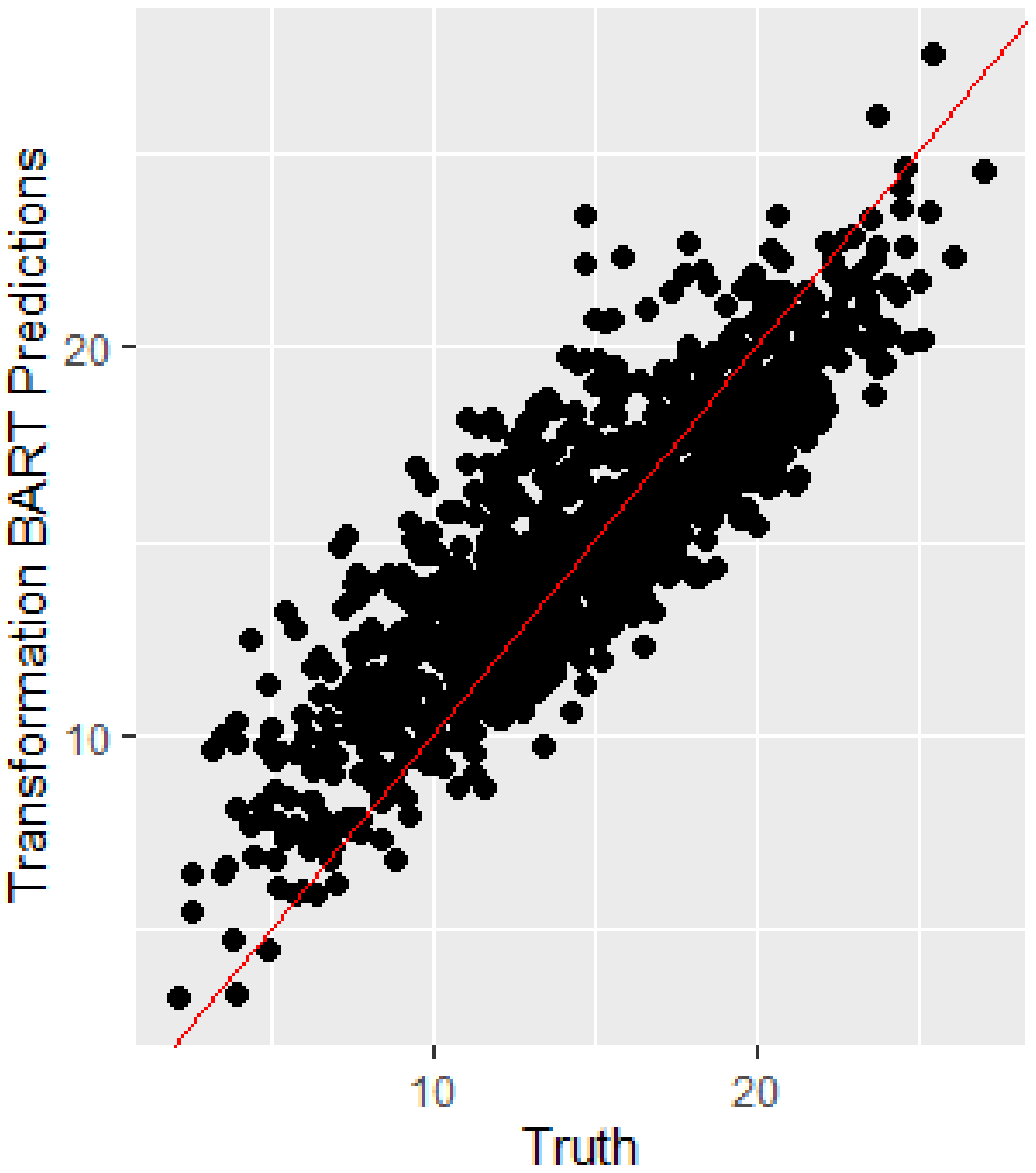}&\includegraphics[width=8cm,height=8cm]{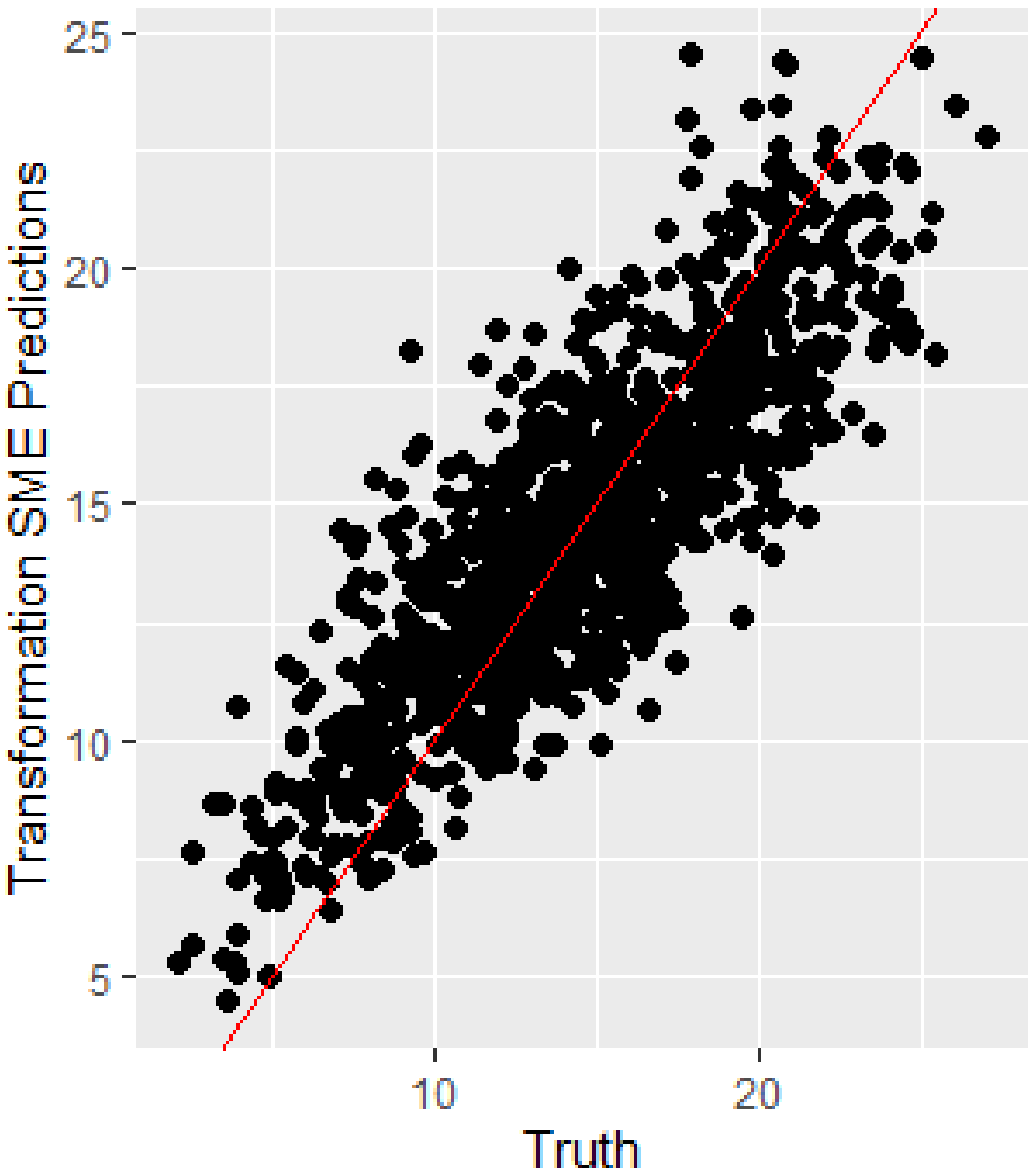}
		\end{tabular}
		\caption{Estimates versus the truth for a single replicate data set. The data are simulated as described in Section~\ref{Section:5.1}. The estimate is labeled on the $y$-axis. The red line indicates the line $y = x$.}\label{fig:2}
	\end{center}
\end{figure}
The data are simulated according to (\ref{simdata}), with $I = 1000$, $I_{1} = 350$, $I_{2} = 350$, and 
$I_{3}=200$. We do not include a validation dataset so that $k_{j}\equiv g_{j}$. We repeat this simulation study 20 times, and we provide violin plots of the RMSE over the 20 replicates by method in Figure \ref{fig:1}. In Figure \ref{fig:2} we also plot the true function versus the estimated function for a single replicate data set. Figures 1 and 2 suggest that the transformation-based spatio-temporal mixed effects (BART) performs well in terms of predictive performance. For the replicate in Figure \ref{fig:2} the transformation-based spatio-temporal mixed effects (and BART) model had $97\%$ ($94\%$) of the point-wise credible intervals of the elements of $\bm{\delta}$ containing zero. The patterns observed in Figure \ref{fig:1} mimic the goodness-of-fit diagnostics, which is notable because the goodness-of-fit diagnostics are data driven (and hence can be used in practice) while Figure \ref{fig:1} is based on the unknown truth.  These results suggests that the Bayesian transformations can be used to obtain predictions in the non-Gaussian setting using two standard models, and also has a useful built-in goodness-of-fit diagnostic. 

\begin{figure}[t]
		\begin{center}
			\begin{tabular}{c}
				\includegraphics[width=10cm,height=8cm]{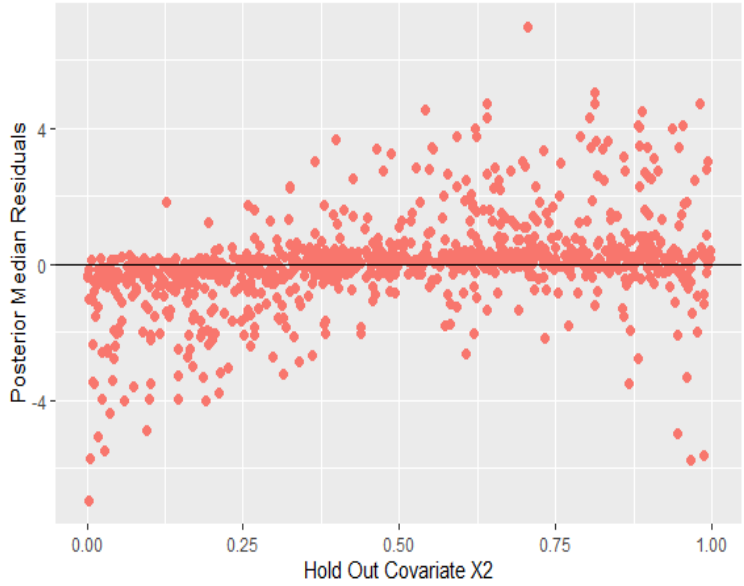}
			\end{tabular}
			\caption{We simulate a single replicate of $\{Y_{ij}\}$ according to Section~\ref{Section:5.1}. Then a spatio-temporal mixed effects model is implemented using the specifications in Section~\ref{Section:3.4}. This plot displays the posterior median of $\{\delta_{ij}\}$ (see Section~\ref{Section:4.1}) versus $\textbf{x}_{2,ij}$, which is not included in our implementation of the spatio-temporal mixed effects model. A systematic pattern in this plot suggests that including $\textbf{x}_{2,ij}$ would improve our analysis of $\textbf{y}$. }\label{fig:res}
		\end{center}
	\end{figure}

Now, suppose we have observed the values of $\{x_{2,ij}\}$, and recall these covariates are not included in the analysis. In Figure \ref{fig:res}, we plot the posterior median of the residuals versus the covariate $\{x_{2,ij}\}$ across the indexes $i$ and $j$ for a single replicate of the data set. The plot clearly indicates a sinusoidal or possibly quadratic pattern, which suggests that this behavior is not captured in our model for $\textbf{y}$. We know this to be true because $\{x_{2,ij}\}$ is not included in our implementation, but the data was generated using $\{x_{2,ij}\}$. This is an illustration of how our approach provides a Bayesian analog to a graphical technique from classical regression analysis (i.e., systematic patterns in residuals from a multiple regression versus a covariate suggest that the covariate should be included in the analysis).
	
\subsection{Simulations: Robustness to Departures from Model Assumptions} In this simulation study we compare the predictive performance our Bayesian transformation approach to predictions from the preferred model itself. A straightforward way to do this is to restrict ourselves to the continuous data-only setting, in which both modeling paradigms can be implemented. 
The data are simulated according to (\ref{simdata}), with $I_{1} = 800$, $I = 1000$, and $I_{2} = I_{3}=0$. We do not include a validation dataset so that $k_{j}\equiv g_{j}$. 

We repeat this simulation study 20 times, and we provide violin plots of the RMSE over the 20 replicates by method in Figure \ref{fig:3}. In this section, we include an additional predictor: soft BART \citep[SBART;][see Appendix B.4 for more details]{linero2018bayesian}. We again see that the Bayesian transformation versions of BART and spatio-temporal mixed effects outperform the saturated model, with the spatio-temporal mixed effects model clearly outperforming BART. Additionally, the Bayesian transformation version of BART and spatio-temporal mixed effects perform only slightly better than or the same as their non-transformed counterparts. Here we see that SBART performs worse than the saturated model in terms of RMSE. The Bayesian transformation version of SBART does not perform noticeably different than SBART in terms of RMSE. Thus, in the continuous only setting, if the preferred model performs well (or poorly) one should expect the Bayesian transformation approach to perform well (or poorly). Recall that we can use the goodness-of-fit approach in Section~\ref{Section:4.1} to assess when a method performs poorly in practice. For example, for a single replicate data set, we found that the percent of credible intervals of the elements of $\bm{\delta}$ that contain zero (by method) are as follows:  $99.8\%$ (spatio-temporal mixed effects), $77.4\%$ (BART), and $58.1\%$ (SBART). This produces the same rankings of the method in terms of RMSE. 
\begin{figure}[t]
	\begin{center}
		\begin{tabular}{cc}
			\includegraphics[width=8cm,height=8cm]{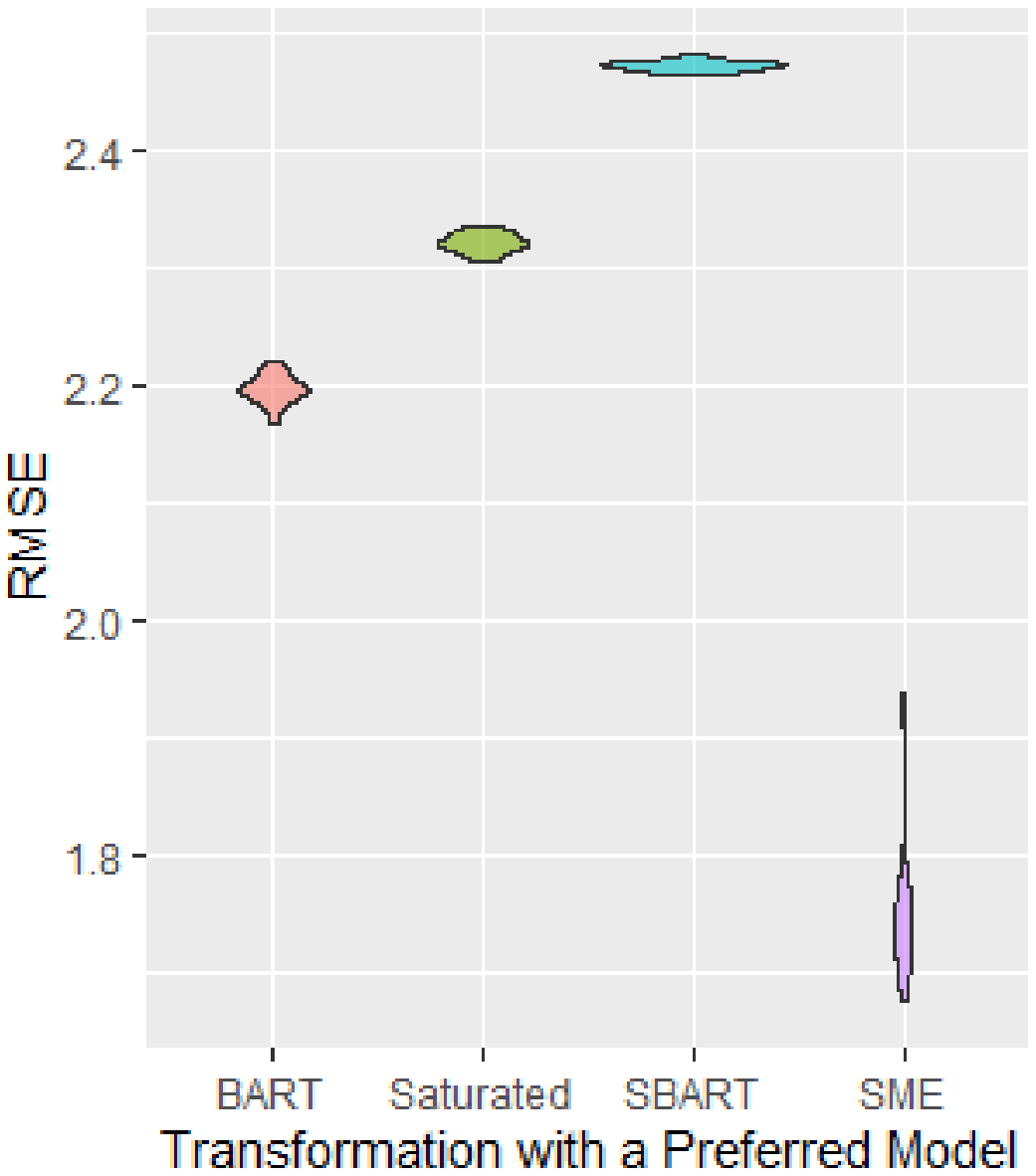}&\includegraphics[width=8cm,height=8cm]{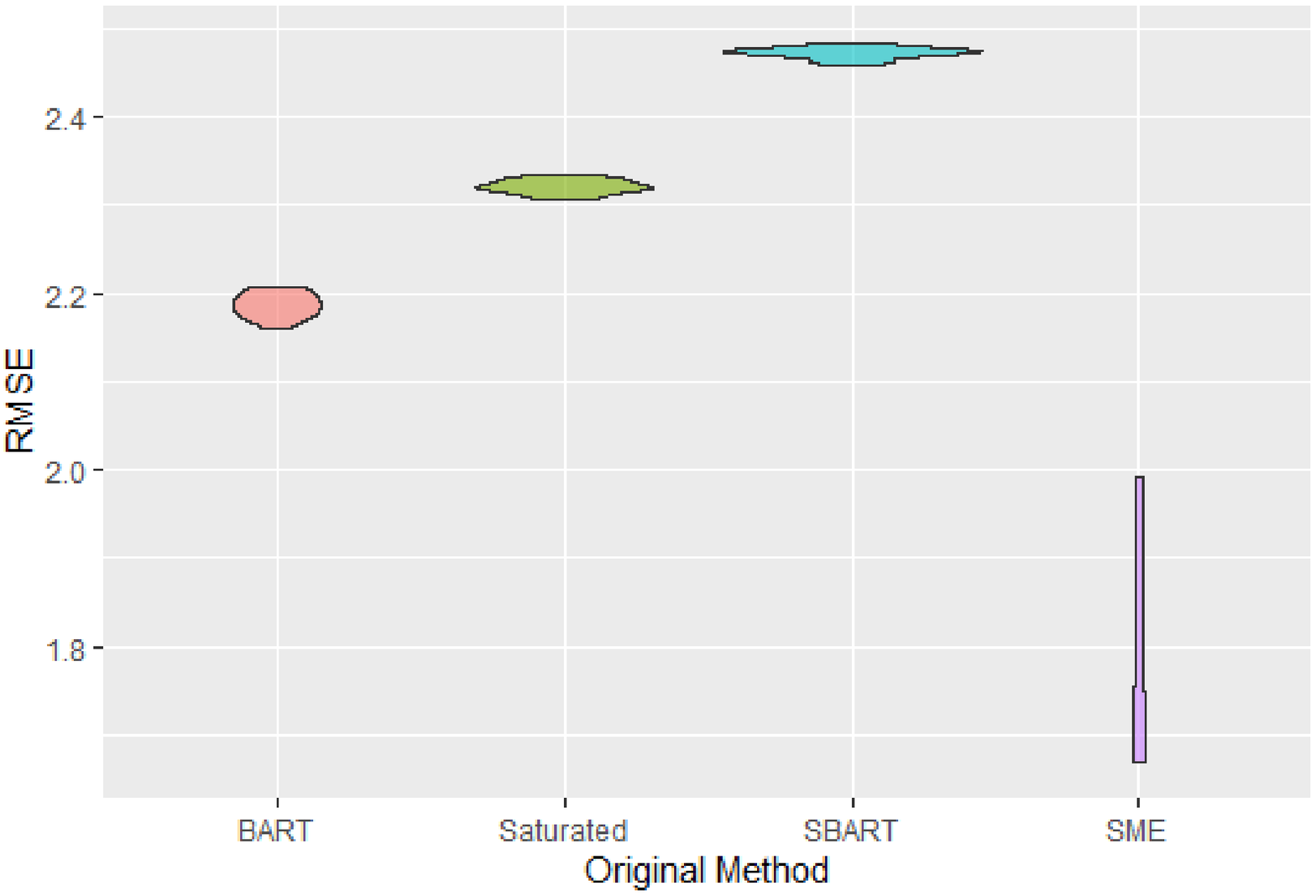}
		\end{tabular}
		\caption{A violin plot of the RMSE (y-axis) by method (x-axis) over 20 independent replicates of the data. The data are simulated as described in Section~\ref{Section:5.1}. Each method is implemented using Algorithm 1, except the method ``Saturated.'' The observed data set are used as the predicted values for the method ``Saturated.'' The left panel displays the results of the Bayesian transformation methods, and the right panel presents the results of the original methods.}\label{fig:3}
	\end{center}
\end{figure}

\begin{figure}[htp]
	\begin{center}
		\begin{tabular}{ccc}
			\hspace{-40pt}	\includegraphics[width=6cm,height=6cm]{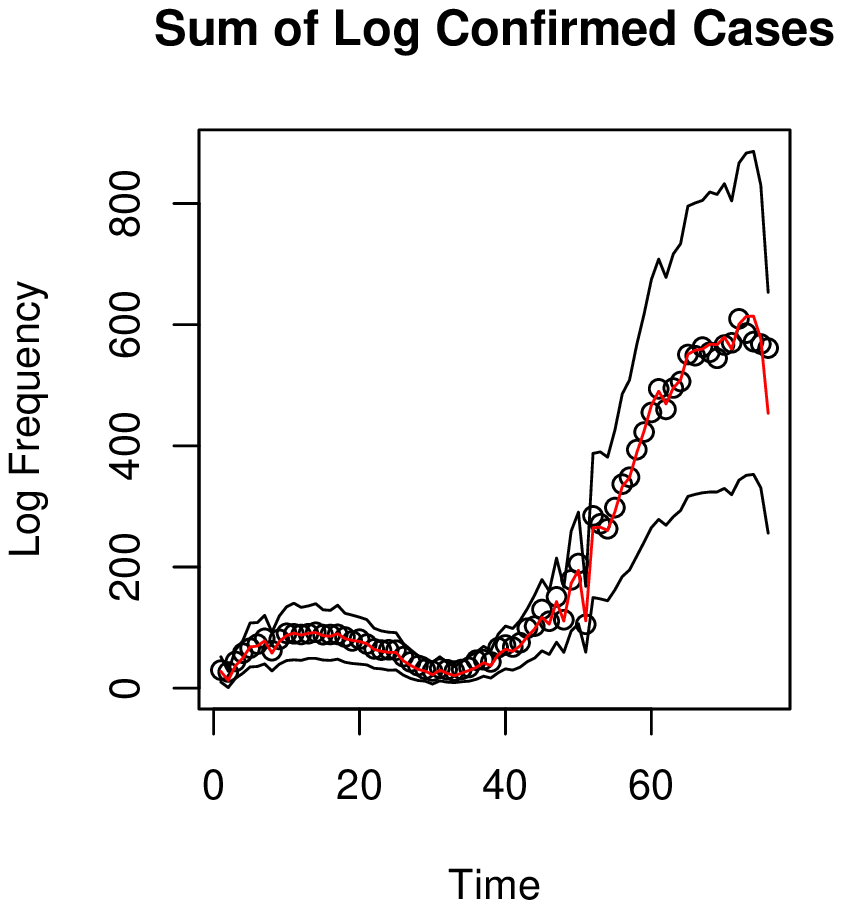}&\includegraphics[width=6cm,height=6cm]{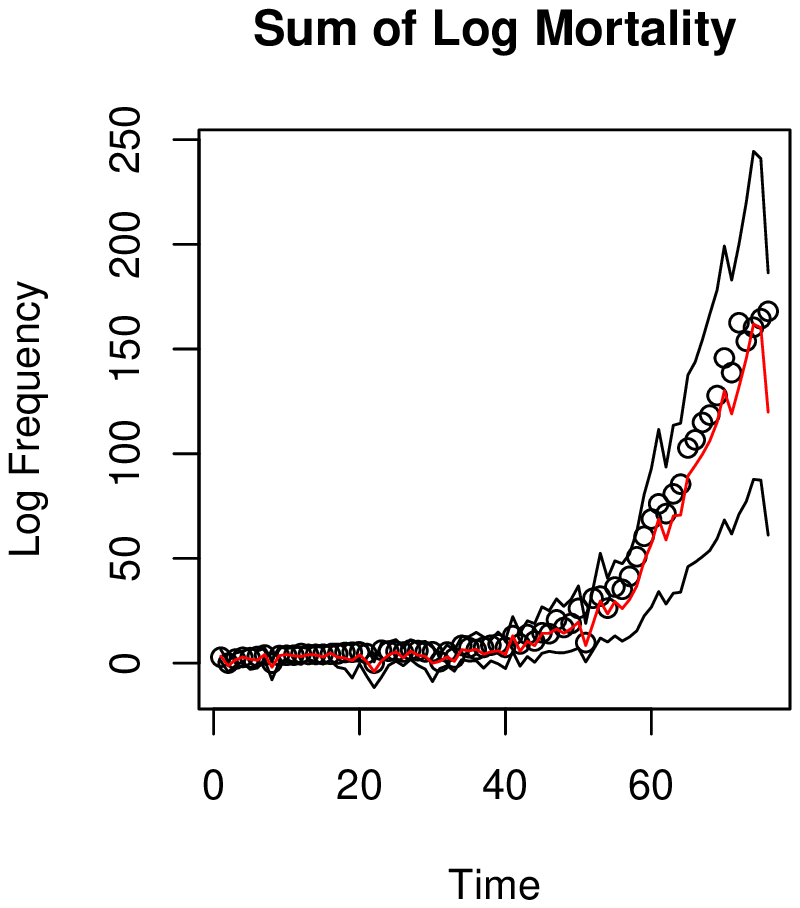}&\includegraphics[width=6cm,height=6cm]{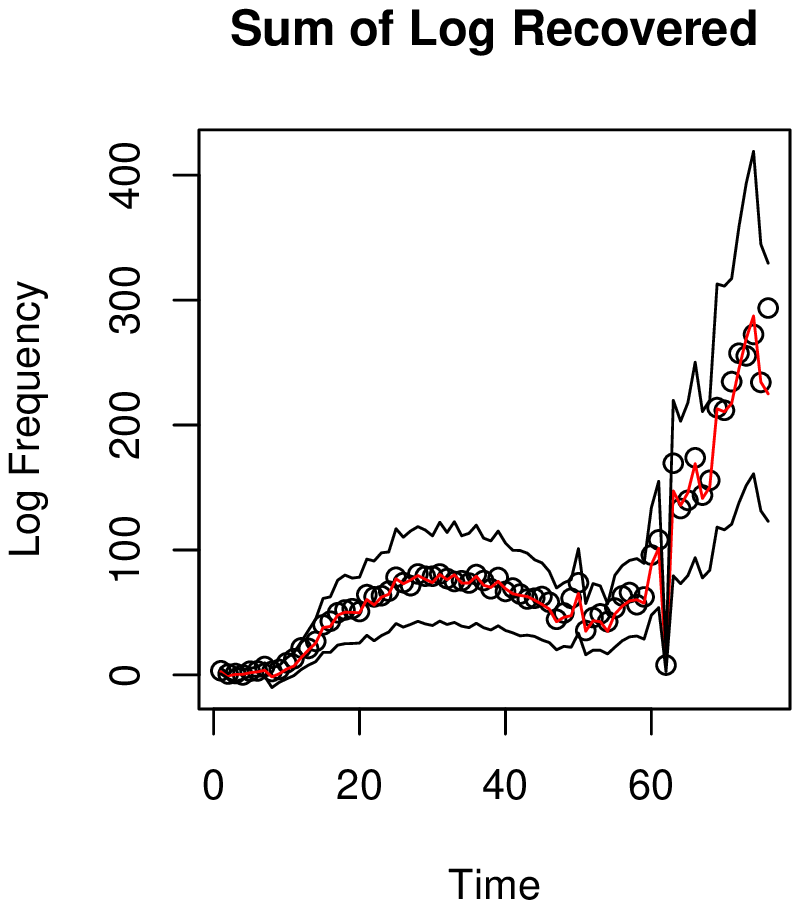}
		\end{tabular}
		\end{center}
	\begin{center}
		\begin{tabular}{ccc}
			\hspace{-40pt}
			\includegraphics[width=6cm,height=6cm]{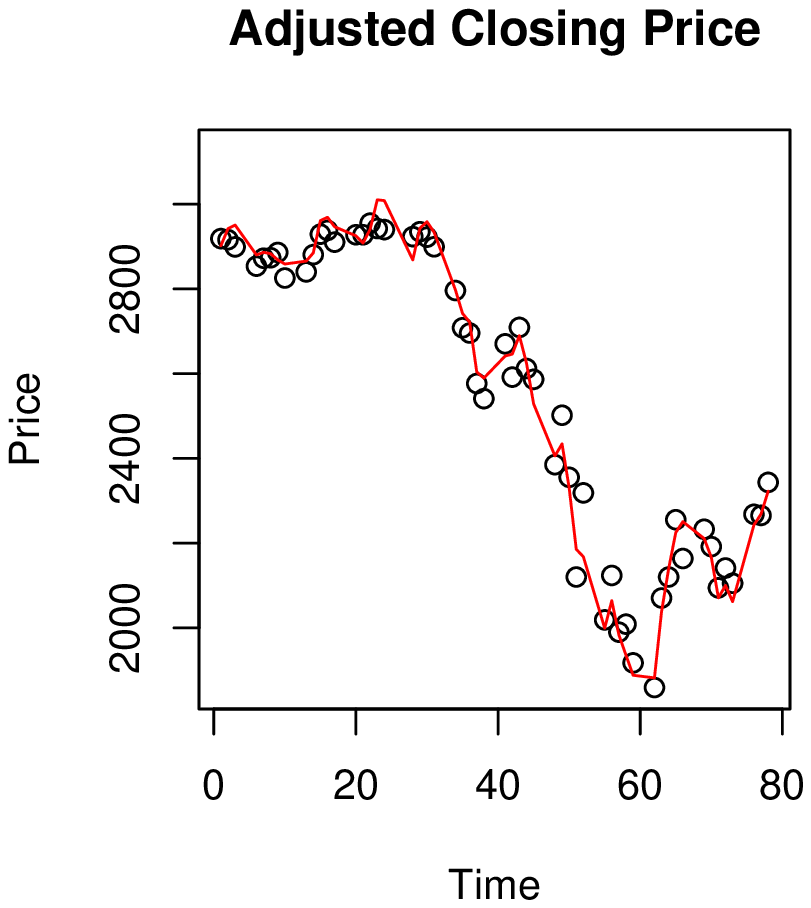}&\includegraphics[width=6cm,height=6cm]{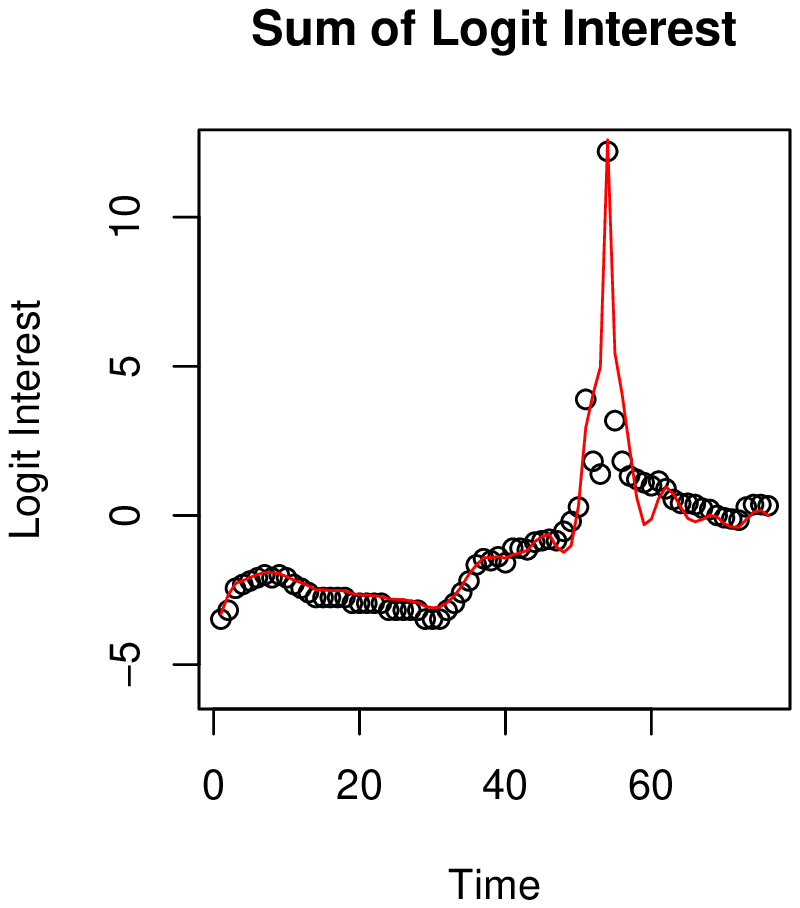}&\includegraphics[width=6cm,height=6cm]{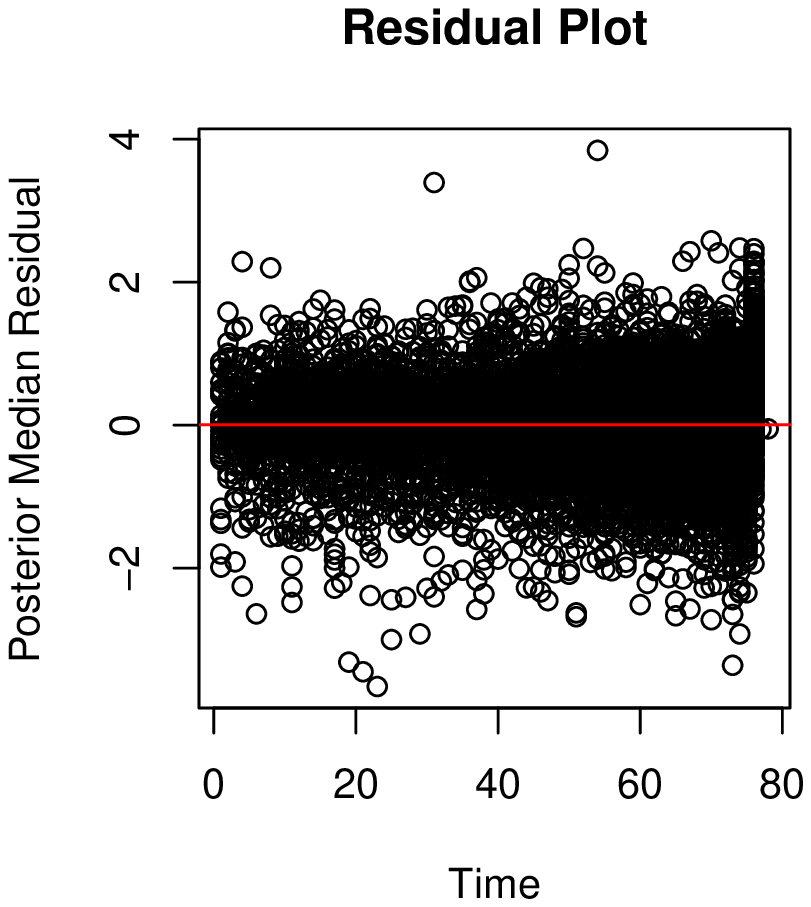}\\
		\end{tabular}
		\caption{Goodness of Fit: We plot the sum (over regions) of log number of reported COVID-19 infections (top left), sum (over regions) log number of reported COVID-19 deaths (top middle), sum (over regions) log number of reported COVID-19 recoveries (top right), the DJI adjusted closing price (bottom left), and the logit ($log(Y_{i2}/100-Y_{i2})$) Google Trends interest score for searches of ``coronavirus'' (bottom middle). Note that the DJI price data is not available on Saturday and Sundays. The red lines represent the predicted values from our model, and the black circle represent the observed values. The black lines are pointwise 95$\%$ credible intervals. The credible intervals are left out in the bottom panels for visualization purposes (credible intervals are large), and in this panel each datum falls within their respective credible interval. The posterior median residuals versus time is given in the bottom right panel.}\label{fig:preds}
	\end{center}
\end{figure}
 \section{Joint analysis of COVID-19 occurrences, the adjusted closing price of the Dow Jones Industrial, and Google Trends data} \label{Section:6}

We now present our joint analysis of deaths due by and occurrences of COVID-19, the adjusted closing price of the DJI, and the Google Trends interest score in searches of ``coronavirus'' (see time series displays of this data in Figure 1). We implement the HGT model, and assume the process and priors in (\ref{summary33}). In our model $Z_{i1}$ represents the negative adjusted closing price per $\$$10,000. This transformation is made so that we see increasing trend over time among all three response types. Our specifications of the basis functions are defined in Appendix B.2, and covariates for the region and response-type are included. The data from January 22, 2020 to April 6, 2020 are the training data ($n=10,600$), the data on April 7, 2020 is held-out as a validation dataset (373 observations), and the data on April 8 is held-out as a testing dataset (374 observations).

The MCMC is implemented according to Algorithms 1 through 3 with 10,000 replicates and a burn-in of 1,000. Convergence was assessed visually through the use of trace plots and through Gelman-Rubin diagnostics \citep{gelmanbook} with no indications of a lack of convergence. All of our analyses were implemented on Windows 10 with the following specifications: Intel(R) CORE(TM) i5-8250U CPU with 1.60Gh.

\subsection{Goodness of Fit} In Figure \ref{fig:preds} we plot the posterior mean death, confirmed cases, recovered cases, adjusted closing price, and Google Trends interest score. Here, we see that the predicted values are reasonably close to their observed values with the observed data close contained within a pointwise 95$\%$ credible interval. These results suggest that the in-sample error is small, and that the predicted values reflect the general patterns of the data. Goodness of fit can be formally investigated according to Section~\ref{Section:4.1}. Roughly 99.4$\%$ percent of the credible intervals, as defined in (\ref{credible_intervals}), contain zero. This provides additional evidence the model provides a reasonable fit to the data. In the bottom right panel of Figure \ref{fig:preds} we plot the posterior median residual (i.e., $\bm{\delta}$) versus the time the observation was recorded. Here we see roughly no pattern over time, which suggests that our specification of the basis functions were reasonable.
\begin{figure}[t]
	\begin{center}
		\begin{tabular}{cc}
			\includegraphics[width=8cm,height=8cm]{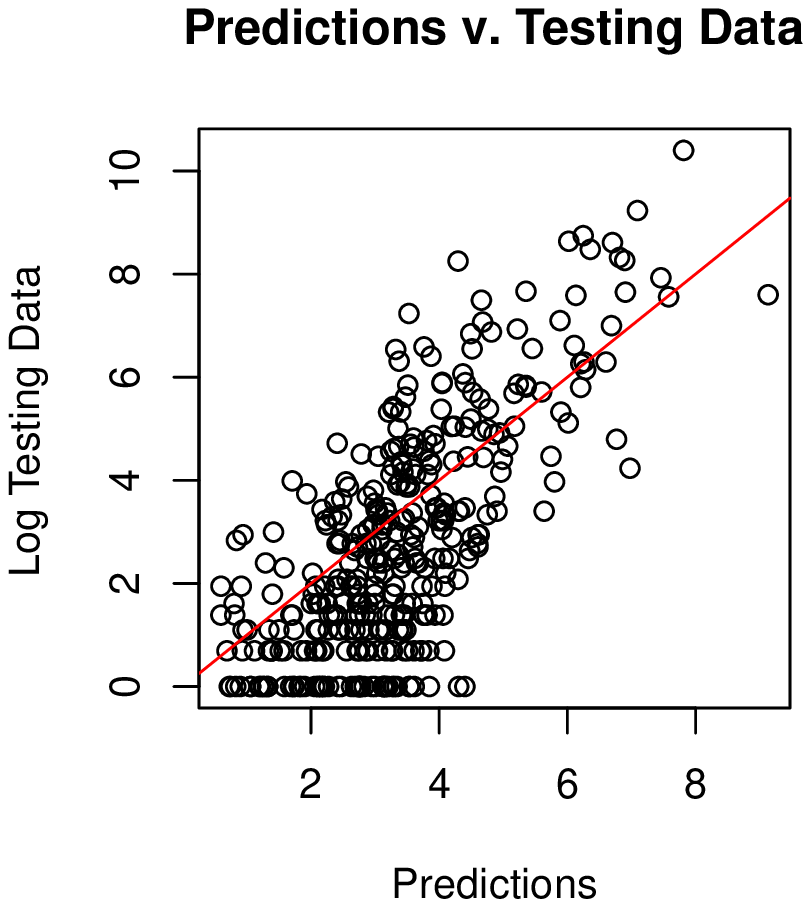}&\includegraphics[width=8cm,height=8cm]{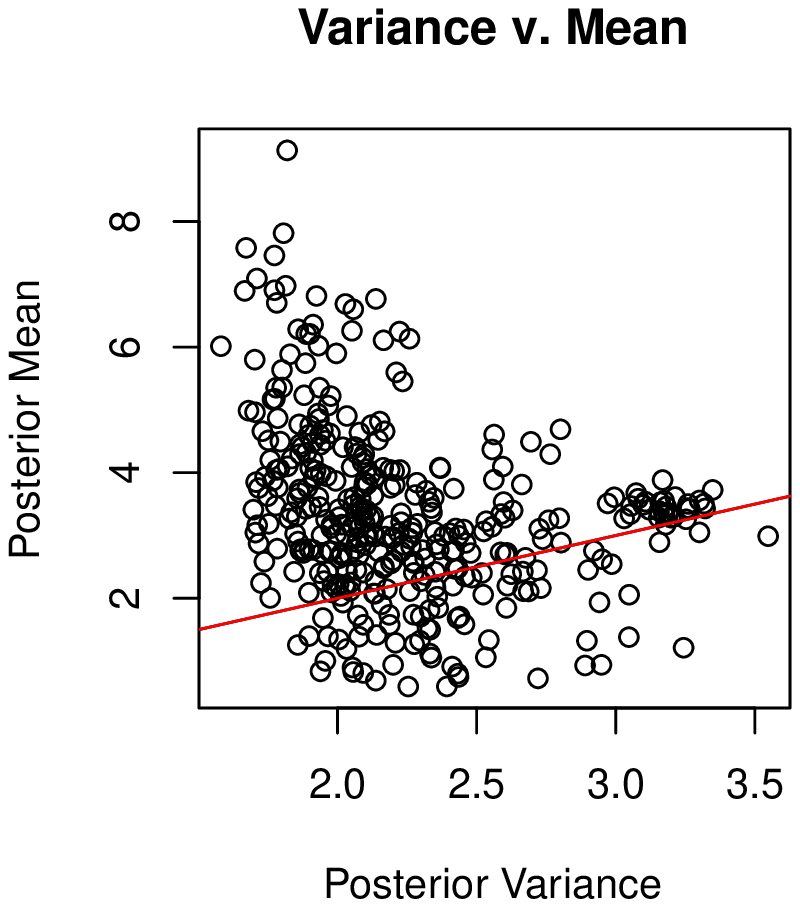}\\
		\end{tabular}
		\caption{Forecasting: In the left panel we plot the forecasted testing data using Algorithm 3. Here the testing data represents all confirmed cases, recoveries, and deaths on April 8, 2020. The right panel plots the posterior variance of the predicted testing data versus the posterior mean.}\label{fig:preds2}
	\end{center}
\end{figure}

\subsection{Estimation and Prediction} We did not include the data on April 8-th, 2020, which was the most current value available at the time of the analysis. We use the model to predict the number of deaths, number of confirmed recoveries, and number of confirmed cases according to Algorithm 3. In Figure~\ref{fig:preds2} we provide the posterior means associated with these values versus the testing data. In general, the posterior means trends the testing data, except for smaller testing values, where there is a tendency to overestimate the log count. However, the percentage (over the testing data) of pointwise credible intervals that contain the the testing data is 98.4$\%$, which suggest that the uncertainty of these estimates are captured in the model. This property of the model is also seen in the plot of the posterior variance versus the posterior mean, also displayed in Figure~\ref{fig:preds2}. Here, smaller predicted values tend to be over-dispersed, and larger predicted values appear to be equi-dispersed. Thus, we appear to have accurate predictions of the areas with the largest confirmed cases, recoveries, and deaths. Being able to accurately estimate large values of (log) occurrences is particularly important. That is, if we know \textit{where} there are large occurrences of confirmed cases, then additional testing of individuals in these regions allows one to isolate all those who test positive in this region, which ultimately reduces the spread of COVID-19 from this region to others \citep{ai2020correlation}. Consequently, models such as ours can be useful at stopping the spread of COVID-19. However, finer-scale regional data would be necessary for this model to be helpful in narrowing in on potential ``hot-spots'' in practice.

\begin{figure}[t]
	\begin{center}
		\begin{tabular}{c}
			\includegraphics[width=10cm,height=10cm]{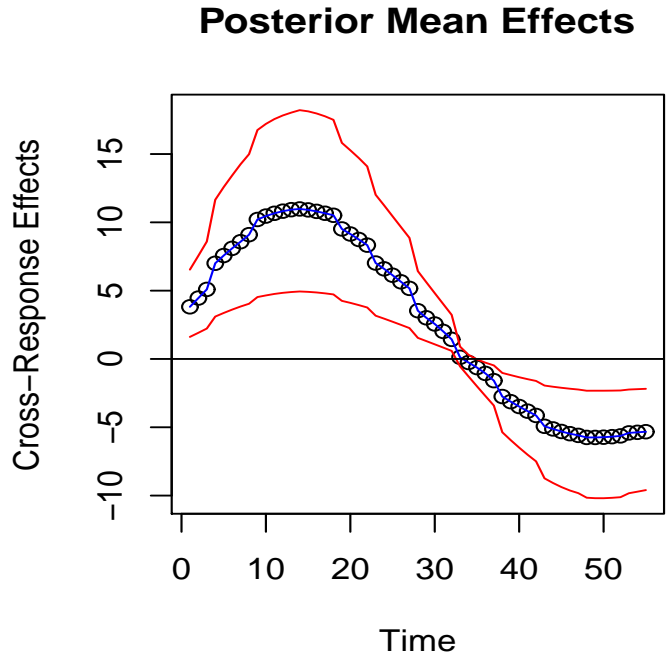}\\
		\end{tabular}
		\caption{We plot the posterior mean of $\sum_{T_{i}=t}\textbf{S}_{ij}^{\prime}\bm{\eta}$. The red line indicates pointwise 95$\%$ credible intervals.}\label{fig:effects}
	\end{center}
\end{figure}

In Figure~\ref{fig:effects}, we plot the posterior mean of the random effects that is shared across response-type along with pointwise 95$\%$ credible intervals (see Section~\ref{Section:4.1}).  The time period between January 22, 2020 and February 23, 2020 was particularly crucial, since this time range saw the strongest direct effects between between COVID-19 cases, the negative adjusted closing price, and Google Trends interest-score in the Google search ``coronavirus.'' Furthermore, the fact that zero does not tend to fall within the credible intervals suggests that our incorporation of dependence across response-types, spatial regions, and days was reasonable. February 23, 2020 ($t_{i} = 33$) marks the time in which the adjusted closing price initially started to decrease (see Figures 1 and 6), and the Google Trends interest score increases. After February 23, 2020 the random effect appears to be negative-valued, which suggests an indirect relationship among these responses. 

\section{Discussion} \label{Section:7}
COVID-19 is a global epochal health disaster, and social distancing has become a necessary public health measure to protect the health of individuals. In this article, we investigate the relationship between COVID-19 cases, the US economy (specifically the adjusted closing price of DJI), and interest on Google (specifically Google Trends interest score for the search ``coronavirus''). The data and model suggests that the relationship among these three values had the strongest positive relationship during a majority of February 2020, which suggests that this was an important time period. Additionally, there are clear cross-dependencies among response types, regions, and days. It is important to comment that correlation does not imply causation, and to make explicit causal conclusions one needs to adopt methods among the causal inference literature \citep{rubin2005causal}. Finally, our model produces reasonable forecasts of the log frequency of cases, deaths, and recoveries from COVID-19. This suggests that with finer-scale regional data, this model could potentially be useful for targeting future hot-spots of COVID-19. 

We introduce the HGT model in order to analyze COVID-19 and social distancing related variables, which is derived from a straightforward combination of the LCM and the GLMM. This combination is motivated as a means to aid other researchers to analyze multi-response datasets such as the one considered in this article. In particular, our approach provides several contributions to Bayesian statistics. First, we have developed a general all-purpose Bayesian model to analyze multiple responses (e.g., continuous, Binomial counts, and Poisson counts). Our approach allows one to  directly incorporate their preferred Bayesian model to analyze multi-response data without completely abandoning their approach to the implementation of their preferred model. Second, we developed a general Bayesian analog to the classical comparison between a saturated model and a reduced model. This results in the use of classical residual analysis for assessing goodness-of-fit in Bayesian models for multi-response data. Code and tutorials on how to adapt the HGT to your preferred model can be found at \url{https://github.com/JonathanBradley28/CM}.

In our simulations, an illustration was given of non-linear functional analysis of multiple response types using BART as the preferred model. Additionally, an illustration was given of a joint spatial analysis of multiple response types using a spatio-temporal mixed effects model as the preferred model. These results suggest that the prediction error of our approach is small (in terms of RMSE), and we can develop multi-response versions of two different preferred models seamlessly. Additionally, data driven goodness-of-fit diagnostics were able to lead to the same conclusion as the RMSE criterion (based on the latent process) that is unobserved in practice.

\section*{Acknowledgments} This research was partially supported by the U.S. National Science Foundation (NSF) under NSF grant SES-1853099. I also would like to thank Drs. Christopher Wikle and Scott Holan at the University of Missouri on their feedback on an earlier version of this article.
%
  
  \singlespacing
\bibliographystyle{jasa} 
\bibliography{myref}

\begin{thebibliography}{71}
\bibitem[\protect\citename{Bradley et~al.,
  }2019{\natexlab{b}}]{bradleyhierarchical}
Bradley, J.~R., Wikle, C.~K., and Holan, S.~H. (2019{\natexlab{b}}).
\newblock \enquote{Hierarchical Models for Spatial Data with Errors that are
  Correlated with the Latent Process.}
\newblock {\em Statistica Sinica\/}.
\end{thebibliography}

\doublespacing

\section*{Appendix A: Derivations}
\renewcommand{\theequation}{A.\arabic{equation}}
\setcounter{equation}{0}
\textit{Derivation of (\ref{bayesconverter})}:  The distributions in (\ref{preferredModel}) and (\ref{posteriorTrans}) can be used to produce the following expression of the joint distribution of the data, process, and parameters
\begin{equation*}
f(\textbf{z}_{trn},\textbf{y},\bm{\theta}) = \int\int f(\textbf{z}_{trn}\vert \textbf{h})f(\textbf{y},\bm{\theta}\vert \textbf{h}) f(\textbf{h}\vert \bm{\gamma}) f(\bm{\gamma})\hspace{2pt}d\textbf{h}\hspace{2pt}d\bm{\gamma} = \int f(\textbf{y},\bm{\theta}\vert \textbf{h}) f(\textbf{z}_{trn},\textbf{h})\hspace{2pt}d\textbf{h}\hspace{2pt},
\end{equation*}
\noindent
where $f(\textbf{z}_{trn},\textbf{h}) = \int f(\textbf{z}_{trn}\vert \textbf{h}) f(\textbf{h}\vert \bm{\gamma}) f(\bm{\gamma})\hspace{2pt}d\bm{\gamma}$ and we have used the assumption of conditional independence between $\bz$ and $(\by,\bm{\theta})$ given $\textbf{h}$.  Then dividing by $f(\textbf{z}_{trn}) = \int\int f(\textbf{z}_{trn}\vert \textbf{h}) f(\textbf{h}\vert \bm{\gamma}) f(\bm{\gamma})\hspace{2pt}d\textbf{h}\hspace{2pt}d\bm{\gamma}$ yields,
\begin{equation*}
f(\by,\bm{\theta}\vert \bz) = \int f(\by,\bm{\theta}\vert \textbf{h})f(\textbf{h}\vert \bz) d\textbf{h},
\end{equation*}
\noindent
which is the desired result.\\

\noindent
\textit{Derivation of (\ref{saturatedModel}):} Versions of this proof can be found in \citet{diaconis} and \citet{bradleyLCM}. The two distributions in (\ref{transat}) associated with $j = 2$ and $j = 3$ are members of the natural exponential family \citep{casellehman}, which are of the form,
\begin{equation*}
\label{EF}
{f(Z_{ij}\vert h_{ij},\alpha_{j},\kappa_{j})} \propto\mathrm{exp}\left\lbrace Z_{ij}h_{ij} - c_{ij}\psi_{j}(h_{ij})\right\rbrace; \hspace{4pt} i = 1,\ldots, I_{j}, j = 2,3,
\end{equation*}
where $c_{i2} = b_{i}$ and $c_{i3} = 1$. Upon multiplying by (\ref{univ_LG}) we have:
\begin{equation*}
{f(h_{ij}\vert Z_{ij},\alpha_{j},\kappa_{j})}\propto \mathrm{exp}\left\lbrace (Z_{ij}+\alpha_{j})h_{ij} - (\kappa_{j}+c_{ij})\psi_{j}(h_{ij})\right\rbrace \propto \mathrm{DY}(\alpha_{j}+Z_{ij}, \kappa_{j}+c_{ij}; \psi_{j}),
\end{equation*}
\noindent
which proves the result for $j=2$ and $j=3$. For $j = 1$,
\begin{align*}
&{f(h_{i1}\vert Z_{i1},\alpha_{1},\kappa_{1})} \propto \mathrm{exp}\left\lbrace \left(\frac{Z_{i1}}{v}+\alpha_{1}\right)h_{i1} - \left(\kappa_{1}+\frac{1}{2v}\right)h_{ij}^{2}\right\rbrace \\
&= \mathrm{exp}\left\lbrace 2\left(2\kappa_{1}+\frac{1}{v}\right)\left(2\kappa_{1}+\frac{1}{v}\right)^{-1}\left(\frac{Z_{i1}}{v}+\alpha_{1}\right)\frac{h_{i1}}{2} - \left(2\kappa_{1}+\frac{1}{v}\right)\frac{h_{ij}^{2}}{2}\right\rbrace \\
&\propto \mathrm{exp}\left\lbrace 2\left(2\kappa_{1}+\frac{1}{v}\right)\left(2\kappa_{1}+\frac{1}{v}\right)^{-1}\left(\frac{Z_{i1}}{v}+\alpha_{1}\right)\frac{h_{i1}}{2} - \left(2\kappa_{1}+\frac{1}{v}\right)\frac{h_{ij}^{2}}{2}\right.\\
&\left. -\frac{1}{2}\left(2\kappa_{1}+\frac{1}{v}\right)\left(2\kappa_{1}+\frac{1}{v}\right)^{-2}\left(2\kappa_{1}+\frac{1}{v}\right)^{-1}\left(\frac{Z_{i1}}{v}+\alpha_{1}\right)^{2} \right\rbrace \\
&=\mathrm{exp}\left[ \frac{\left\lbrace h_{i1} -\left(2\kappa_{1}+\frac{1}{v}\right)^{-1}\left(\frac{Z_{i1}}{v}+\alpha_{1}\right) \right\rbrace^{2}}{2\left(2\kappa_{1}+\frac{1}{v}\right)^{-1}}\right]\\
&\propto \mathrm{Normal}\left\lbrace \left(2\kappa_{1}+\frac{1}{v}\right)^{-1}\left(\frac{Z_{i1}}{v} + \alpha_{1}\right), \left(2\kappa_{1}+\frac{1}{v}\right)^{-1}\right\rbrace,
\end{align*}
\noindent
which completes the results.\\

\noindent
\textit{Derivation of (\ref{overfitit}):} In Equation (\ref{Step2}) we see that 
\begin{equation*}
E(h_{i1}\vert Z_{i1},\bm{\gamma}) = \left(2\kappa_{1} + \frac{1}{v}\right)^{-1}\left(\frac{Z_{i1}}{v}+\alpha_{1}\right) + E(w_{1}\vert Z_{i1},\bm{\gamma}) = \left(2\kappa_{1} + \frac{1}{v}\right)^{-1}\left(\frac{Z_{i1}}{v}+\alpha_{1}\right),
\end{equation*}
\noindent
which converges to $Z_{i1}$ as $\alpha_{1}$ and $\kappa_{1}$ approach zero. The expectation of a beta distribution is well known \citep{casellaBerger}, which from (\ref{Step2}) gives us
\begin{equation*}
E\left\lbrace g(h_{i2})\vert Z_{i2},\bm{\gamma}\right\rbrace =  E(w_{2}\vert Z_{i1},\bm{\gamma}) = \frac{\alpha_{2} + Z_{i2}}{\kappa_{2}+b_{i}},
\end{equation*}
\noindent
which converges to $Z_{i2}/b_{i}$ as $\alpha_{2}$ and $\kappa_{2}$ approach zero. Similarly, the expectation of a gamma distribution is well known \citep{casellaBerger}, which from (\ref{Step2}) gives us
\begin{equation*}
E\left\lbrace g(h_{i3})\vert Z_{i3},\bm{\gamma}\right\rbrace =  E(w_{3}\vert Z_{i1},\bm{\gamma}) = \frac{\alpha_{3} + Z_{i3}}{\kappa_{3}+1},
\end{equation*}
\noindent
which converges to $Z_{i3}$ as $\alpha_{3}$ and $\kappa_{3}$ approach zero.\\

\noindent
\textit{Proof that (\ref{summary4}) is proper:} The joint distribution of the training data, transformed data, process, parameters, and transformation hyperprior is given by:
\begin{equation*}
\left\lbrace\prod_{i = 1}^{I_{1}}f(Z_{i1}\vert h_{i1})\right\rbrace \left\lbrace\prod_{i = 1}^{I_{2}}f(Z_{i2}\vert h_{i2})\right\rbrace \left\lbrace\prod_{i = 1}^{I_{3}}f(Z_{i3}\vert h_{i3})\right\rbrace f(\textbf{h}\vert \textbf{y},\bm{\theta})m(\textbf{h}\vert \bm{\gamma})f(\textbf{y}\vert\bm{\theta}) f(\bm{\theta})f(\bm{\gamma}).
\end{equation*}
Then integrate out $\textbf{y}$ and $\bm{\theta}$ to obtain,
\begin{equation*}
\left\lbrace\prod_{i = 1}^{I_{1}}f(Z_{i1}\vert h_{i1})\right\rbrace \left\lbrace\prod_{i = 1}^{I_{2}}f(Z_{i2}\vert h_{i2})\right\rbrace \left\lbrace\prod_{i = 1}^{I_{3}}f(Z_{i3}\vert h_{i3})\right\rbrace \left\lbrace \prod_{i,j}f_{DY}(h_{ij}\vert \alpha_{j}, \kappa_{j}, a, b)\right\rbrace f(\bm{\gamma}),
\end{equation*}
which follows from,
\begin{align*}
&\int \int f(\textbf{h}\vert \textbf{y},\bm{\theta})m(\textbf{h}\vert \bm{\gamma})f(\textbf{y}\vert\bm{\theta}) f(\bm{\theta}) d\textbf{y}d\bm{\theta} \\
&= \int \int f(\textbf{h}\vert \textbf{y},\bm{\theta})f(\textbf{y}\vert\bm{\theta}) f(\bm{\theta}) d\textbf{y}d\bm{\theta}\frac{\prod_{i,j}f_{DY}(h_{ij}\vert \alpha_{j}, \kappa_{j}, a, b)}{\int\int(f(\textbf{h}\vert \textbf{y}, \bm{\theta})f(\textbf{y}\vert \bm{\theta})f(\bm{\theta})d\textbf{y}d\bm{\theta}} =\prod_{i,j}f_{DY}(h_{ij}\vert \alpha_{j}, \kappa_{j}, a, b).
\end{align*}
\noindent
Finally, we have the result, since the normal, binomial, Poisson, and DY distributions \citep{diaconis} are proper and the prior on $\bm{\gamma}$ is proper.

\section*{Appendix B: Additional Model Details}

\subsection*{Appendix B.1: Full-Conditional Distributions for the Transformation Hyperparameters} 
\renewcommand{\theequation}{B.1.\arabic{equation}}
\setcounter{equation}{0}

The full-conditional distributions for the transformation hyperparameters are found by multiplying $f(\textbf{h}\vert \bm{\gamma})$ and $f(\bm{\gamma})$ as follows:
\begin{align}\label{hyp}
\nonumber
v\vert \cdot &\sim IG\left(\frac{I_{1}}{2}+a_{1},\frac{\sum_{i = 1}^{I_{2}}(Z_{i1}-h_{i1})}{2}+b_{1}\right)\\
\nonumber
f(\alpha_{2}\vert \cdot )&\propto \alpha_{2}^{a_{2}-1}\mathrm{exp}(-b_{2}\alpha_{2})\frac{1}{\Gamma(\alpha_{2})^{I_{2}}\Gamma(\kappa_{2}-\alpha_{2})^{I_{2}}}\mathrm{exp}(\alpha_{2}\sum_{i = 1}^{I_{2}}h_{i2})\\
\nonumber
f(\alpha_{3}\vert \cdot )&\propto \alpha_{3}^{a_{3}-1}\mathrm{exp}(-b_{3}\alpha_{3})\frac{\kappa_{3}^{I_{3}\alpha_{3}}}{\Gamma(\alpha_{3})^{I_{3}}}\mathrm{exp}(\alpha_{3}\sum_{i = 1}^{I_{3}}h_{i3})\\
\nonumber
f(\kappa_{2}\vert \cdot )&\propto (\kappa_{2}-\alpha_{2})^{\zeta_{2}-1}\mathrm{exp}(-\eta_{2}\kappa_{2})\frac{\Gamma(\kappa_{2})^{I_{2}}}{\Gamma(\kappa_{2}-\alpha_{2})^{I_{2}}}\mathrm{exp}(-\kappa_{2}\sum_{i = 1}^{I_{2}}\mathrm{log}(1+\mathrm{exp}(h_{i2})))\mathcal{I}(\kappa_{3}\ge \alpha_{3})\\
f(\kappa_{3}\vert \cdot )&\propto (\kappa_{3}-\alpha_{3})^{\zeta_{3}-1}\mathrm{exp}(-\eta_{3}\kappa_{3})\kappa_{3}^{I_{3}\alpha_{3}}\mathrm{exp}(-\kappa_{3}\sum_{i = 1}^{I_{3}}\mathrm{exp}(h_{i3}))\mathcal{I}(\kappa_{3}\ge \alpha_{3}),
\end{align}
\noindent
where $\Gamma(t) = \int_{0}^{\infty}x^{t-1}\mathrm{exp}(-x)dx$, $\mathcal{I}(\cdot)$ is the indicator function, and $IG(a,b)$ is an inverse gamma distribution with shape $a>0$ and rate $b>0$. In our implementation we set the parameters $a_{1} = a_{2} = a_{3} = \zeta_{2}=\zeta_{3} 1$ and $b_{1} = b_{2}= b_{3} = \eta_{2}=\eta_{3} = 1$. We have found that our results are robust to this specification. Step 3 of Algorithm 1 involves simulating from the full conditional distributions in (\ref{hyp}).

\subsection*{Appendix B.2: Choices of Basis Functions} 
\renewcommand{\theequation}{B.2.\arabic{equation}}
\setcounter{equation}{0}

In Section~\ref{Section:5}, the $r$-dimensional real-valued vector $\textbf{S}_{ij}$ is defined to be the Moran's I basis function \citep{hughes}. The Moran's I basis functions \citep{griffith2000,griffith2002,griffith2004} are motivated as a way to remove confounding between $\bm{\beta}$ and $\bm{\eta}$, and allow for dimension reduction. The basis functions are derived from the Moran's I operator used in spatial statistics \citep{MoranI}. Specifically, basis functions are specified to be in the orthogonal column space associated with the hat matrix $\textbf{X}(\textbf{X}^{\prime}\textbf{X})^{-1}\textbf{X}^{\prime}$, where the $N\times p$ matrix $\textbf{X} = \left(\textbf{x}_{ij}: i = 1,\ldots, I, j = 1,2,3\right)$. Define the Moran's I operator
\begin{equation*}
\textbf{G}(\textbf{X},\textbf{A}_{t}) \equiv\left(\textbf{I}_{N}-\textbf{X}(\textbf{X}^{\prime}\textbf{X})^{-1}\textbf{X}^{\prime}\right)\textbf{W}\left(\textbf{I}_{N}-\textbf{X}(\textbf{X}^{\prime}\textbf{X})^{-1}\textbf{X}^{\prime}\right),
\end{equation*}
where $\textbf{W}$ is a generic real-valued $N\times N$ matrix, which is often specified to be an adjacency matrix that characterizes a network.  The spectral representation $\textbf{G}(\textbf{X},\textbf{W}) = \bm{\Phi}\bm{\Lambda}{\bm{\Phi}^{\prime}}$, is computed using a $N\times N$ orthogonal matrix $\bm{\Phi}$ and a $N\times N$ diagonal matrix with positive elements $\Lambda$. Let the $N\times r$ real matrix consisting of the first $r$ columns of $\bm{\Phi}$ be denoted by $\textbf{S}$. The row of $\textbf{S}$ corresponding to the $(i,j)$-th data is set to equal to $\textbf{S}_{ij}$. In Section~\ref{Section:5}, we set $r = 500$.

In Section~\ref{Section:5}, the $r$-dimensional real-valued vector $\textbf{S}_{ij}$ is defined to be thin-plate splines \citep{wahba}. Specifically, let the $m$-th element of the $10$-dimensional vector $\textbf{S}_{i1}^{(k)}$ be defined as,
\begin{equation}
(t_{i}/78 - c_{m})^{2}log\left\lbrace abs(t_{i}/78 - c_{m})\right\rbrace,
\end{equation}
\noindent
where ${c_{m}} =\{0, 0.11, 0.22, 0.33, 0.44, 0.56, 0.67, 0.78, 0.89, 1\}$ are 10 equally spaced values over $\{t_{1},\ldots, t_{78}\}$. Then, let the $m$-th element of the 25-dimensional vector $\textbf{S}_{ij}^{*}$ be
\begin{equation}
(t_{i}/78 - c_{m}^{*})^{2}log\left\lbrace abs(t_{i}/78 - c_{m}^{*})\right\rbrace,
\end{equation}
\noindent
where $\{c_{m}^{*}\}$ is a set of 25 equally spaced time-points between zero and one. Let the $|A_{k}|\times 10$ matrix $\textbf{S}_{1}^{(k)}= (\textbf{S}_{i1}^{(k)}: A_{i} = A_{k})$ and the $I_{1}\times 2660$ matrix $\textbf{S}_{1}= blkdiag(\textbf{S}_{1}^{(1)}, \ldots, \textbf{S}_{1}^{(266)})$, where $blkdiag$ is the block-diagonal operator and $|A_{k}|$ is the number of observations recorded in region $A_{k}$ so that $I_{1} = \sum_{k}|A_{k}|$. Here,  the $I_{1}\times 2660$ matrix $\textbf{S}_{1}$ defines a set of basis matrices for each of the 266 regions in the study, and hence, we allow for different time series within each region. Note that some regions contain others (e.g., provinces are contained with countries). As such, shared time series within a country imply within-country spatial dependence. Define the $I_{j}\times 25$ matrix $\textbf{S}_{j} = (\textbf{S}_{ij}^{*}: i = 1,\ldots, I_{j})$ for $j = 2, 3$, which defines basis matrices for each individual response types. Then collect all individual-level basis matrices into the matrix $n\times 2710$ matrix $\textbf{S}^{**} = blkdiag(\textbf{S}_{1}, \textbf{S}_{2}, \textbf{S}_{3})$. Let the $n\times 25$ matrix $\textbf{S}^{*} = (\textbf{S}_{ij}^{*}: i = 1,\ldots, I_{j}, j = 1,2,3)$, which represents the set of basis functions that are shared among all response types. Finally, the $n\times 2735$ matrix $\textbf{S} = (\textbf{S}^{*}, \textbf{S}^{**})$ represents the basis matrix used in our analysis, and the 2735-dimensional $(i,j)$-th row is denoted with $\textbf{S}_{ij}$.

\subsection*{Appendix B.3: Full-Conditional Distributions for the Spatio-Temporal Mixed Effects Model} 
\renewcommand{\theequation}{B.3.\arabic{equation}}
\setcounter{equation}{0}
The full conditional distributions for this spatio-temporal mixed effects model are well-known \citep[e.g.,see][for a standard reference reference]{cressie-wikle-book}  and are as follows: 
\begin{align}
\nonumber
\bm{\beta} \vert \cdot&\sim \mathrm{Normal}\left(\bm{\mu}_{\beta}^{*},\bm{\Sigma}_{\beta}^{*}\right); \hspace{15pt}\bm{\mu}_{\beta}^{*} \equiv \frac{1}{\sigma^{2}}\bm{\Sigma}_{\beta}^{*}(\textbf{h}-\bm{\xi}-\textbf{S}\bm{\eta}),\hspace{15pt}\bm{\Sigma}_{\beta}^{*} \equiv
\left(\frac{1}{\sigma^{2}}\textbf{X}^{\prime}\textbf{X} + \frac{1}{\sigma_{\beta}^{2}}\textbf{I}_{p}\right)^{-1},\\
\nonumber
\bm{\eta} \vert \cdot&\sim \mathrm{Normal}\left(\bm{\mu}_{\eta}^{*},\bm{\Sigma}_{\eta}^{*}\right);\hspace{15pt}\bm{\mu}_{\eta}^{*} \equiv \frac{1}{\sigma^{2}}\bm{\Sigma}_{\eta}^{*}(\textbf{h}-\textbf{X}\bm{\beta}-\bm{\xi}), \hspace{15pt}\bm{\Sigma}_{\eta}^{*} \equiv \left(\frac{1}{\sigma^{2}}\textbf{I}_{r} + \frac{1}{\sigma_{\eta}^{2}}\textbf{I}_{r}\right)^{-1}\\
\label{effects}
\bm{\xi}\vert \cdot &\sim \mathrm{Normal}\left(\bm{\mu}_{\xi}^{*},\bm{\Sigma}_{\xi}^{*}\right); \hspace{15pt}\bm{\mu}_{\xi}^{*} \equiv \frac{1}{\sigma^{2}}\bm{\Sigma}_{\xi}^{*}(\textbf{h}-\textbf{X}\bm{\beta}-\textbf{S}\bm{\eta}), \hspace{15pt}\bm{\Sigma}_{\xi}^{*} \equiv \left(\frac{1}{\sigma^{2}}\textbf{I}_{n} + \frac{1}{\sigma_{\xi}^{2}}\textbf{I}_{n}\right)^{-1}.
\end{align}
\noindent
The full conditional distributions for variance parameters are well-known \citep[e.g.,see][for a standard reference]{gelmanbook} and are as follows: 
\begin{align}\label{vars}
\nonumber
\sigma^{2}\vert \cdot &\sim IG\left(\frac{n}{2}+\alpha_{v},\frac{\sum_{i = 1}^{I_{2}}\sum_{j = 1}^{3}(h_{i1}-\textbf{x}_{ij}^{\prime}\bm{\beta}-\textbf{S}_{ij}^{\prime}\bm{\eta}-\xi_{ij})}{2}+\beta_{v}\right)\\
\nonumber
\sigma_{\eta}^{2}\vert \cdot &\sim IG\left(\frac{r}{2}+\alpha_{\eta},\frac{\bm{\eta}^{\prime}\bm{\eta}}{2}+\beta_{\eta}\right)\\
\sigma_{\xi}^{2}\vert \cdot &\sim IG\left(\frac{n}{2}+\alpha_{\xi},\frac{\bm{\xi}^{\prime}\bm{\xi}}{2}+\beta_{\xi}\right).
\end{align}
Step 4 of Algorithm 1 for this model involves simulating from the full-conditional distributions in (\ref{effects}) and (\ref{vars}).

\subsection*{Appendix B.4: Bayesian Additive Regression Trees}\renewcommand{\theequation}{B.4.\arabic{equation}}
\setcounter{equation}{0}
Consider the following expression for the BART model \citep[e.g., see][among others]{chipman2010bart}:
\begin{align}\label{summary555}
\nonumber
&\mathrm{Data\hspace{5pt}Model:}\hspace{5pt}h_{ij}\vert \textbf{M}_{k}, \textbf{T}_{k}, \sigma^{2},\bm{\lambda}\ind \mathrm{Normal}\left\lbrace \sum_{k = 1}^{m}w(\textbf{x}_{ij}; \textbf{M}_{k}, \textbf{T}_{k}), \sigma^{2}\right\rbrace \hspace{5pt} m(\textbf{h}\vert \bm{\lambda});\\
\nonumber
&\mathrm{Prior\hspace{5pt}1:}\hspace{5pt} \mu_{gh}\vert \textbf{T}_{k} \sim \mathrm{Normal}\left(0, \frac{1}{4\epsilon^{2}m}\right);\\
\nonumber
&\mathrm{Prior\hspace{5pt}2:}\hspace{5pt} \sigma^{2} \sim \mathrm{IG}\left(\alpha_{v}, \beta_{v}\right);\\
&\mathrm{Prior\hspace{5pt}3:}\hspace{5pt} f(\textbf{T}_{k})\propto \prod_{g = 1}^{u_{k}}\alpha(1+d_{g})^{-\beta};\hspace{2pt}i = 1,\ldots I_{j}, j = 1,2,3,
\end{align}
\noindent
where $\textbf{x}_{ij}$ is a $p$-dimensional vector of known covariates, $w(\cdot)$ is a decision tree \citep[see definition in][]{chipman2010bart}, set $\textbf{M}_{k} = (\mu_{11} … \mu_{b_{k}k)}^{\prime}$, $b_{k}$ is the $k$-th terminal node, and $d_{k}$ is the depth of internal node ${k}$. The hyperparameters $\epsilon \in [1,3]$, $\alpha_{\nu}>0$, $\beta_{v}>0$, $\alpha>0$, and $\beta>0$ are chosen based on the default specifications of the R package \texttt{BayesTree} \citep{BayesTree}. Implementation is achieved through a Metropolis-within-Gibbs sampler and a backfitting algorithm as described in \citet{chipman2010bart}. This Markov chain Monte Carlo (MCMC) algorithm is computed using the R package \texttt{BayesTree}. That is, Step 4 of Algorithm 1 for this model involves simulating from posterior distribution of $\{\textbf{M}_{k}\}$, $\{\textbf{T}_{k}\}$, and $\sigma^{2}$ using \texttt{BayesTree}. The SBART method is an extension of the BART algorithm, which involves a different specification of $w(\cdot)$. Public use code described in \citet{linero2018bayesian} is used.

\end{document}